\documentclass[prx,reprint,onecolumn,superscriptaddress,showpacs,floatfix]{revtex4-1}
\usepackage{amsmath}
\usepackage{changes}
\usepackage{mathrsfs}
\usepackage{txfonts}
\usepackage[perpage]{footmisc}
\usepackage{amssymb}
\usepackage{color}
\usepackage{graphicx}
\usepackage{hyperref}
\usepackage{ulem}
\usepackage{overpic}
\usepackage{psfrag}
\usepackage{tabularx}
\usepackage{array}
\usepackage{placeins}
\newcommand{\PreserveBackslash}[1]{\let\temp=\\#1\let\\=\temp}
\onecolumngrid

\begin{document}

\title{Goldstone modes and the golden spiral in the ferromagnetic\\ spin-1 biquadratic model}

\author{Huan-Qiang Zhou}\thanks{E-mail: hqzhou@cqu.edu.cn}
\affiliation{Centre for Modern Physics,	Chongqing University, Chongqing 400044, The People's Republic of China}

\author{Qian-Qian Shi}
\affiliation{Centre for Modern Physics,	Chongqing University, Chongqing 400044, The People's Republic of China}

\author{Ian P. McCulloch}
\affiliation{Department of Physics, National Tsing Hua University, Hsinchu 30013, Taiwan}
\affiliation{Frontier Center for Theory and Computation, National Tsing Hua University, Hsinchu 30013, Taiwan}
\affiliation{Centre for Modern Physics, Chongqing University, Chongqing 400044, The People's Republic of China}

\author{Murray T. Batchelor}
\affiliation{Mathematical Sciences Institute, The Australian National University, Canberra ACT 2601, Australia}
\affiliation{Centre for Modern Physics, Chongqing University, Chongqing 400044, The People's Republic of China}

\begin{abstract}
Ferromagnetic ground states have often been overlooked in comparison to seemingly more interesting antiferromagnetic ground states. However, both the physical and mathematical structure of ferromagnetic ground states are particularly rich. We show that the highly degenerate and highly entangled  ground states of the ferromagnetic spin-1 biquadratic model are scale invariant, originating from spontaneous symmetry breaking  from ${\rm SU}(3)$ to ${\rm U}(1)\times {\rm U}(1)$ with two type-B Goldstone modes if the system size is even or from  ${\rm SU}(2)$ to ${\rm U}(1)$ with one type-B Goldstone mode if the system size is odd, when  periodic boundary conditions are adopted. The ground state degeneracies are characterized as Fibonacci-Lucas sequences, under open and periodic boundary conditions, with nonzero residual entropy per site. This implies that 
the ground state degeneracies for this model are asymptotically the golden spiral. In addition, sequences of atypical (periodic) degenerate  ground states generated from highest and generalized highest weight states are constructed to establish that the entanglement entropy scales logarithmically with the block size in the thermodynamic limit. The prefactor is half the number of type-B Goldstone modes, which is identified to be the fractal dimension, if one is restricted to atypical degenerate ground states. We also argue that the same conclusion is valid for typical (non-periodic) degenerate ground states, as long as the block size is sufficiently large.
\end{abstract}

\maketitle

\section{Introduction}
\label{intro}

As a basic notion, spontaneous symmetry breaking (SSB) plays a prominent role in atomic and particle physics, condensed matter physics, astrophysics and cosmology~\cite{andersonbook,SSBbook}.
In particular, if a broken symmetry group is continuous,  a gapless Goldstone mode (GM) emerges.
According to Goldstone~\cite{goldstone,goldstone2,goldstone3}, the number of GMs is equal to the number of broken symmetry generators $N_{BG}$ for a relativistic system undergoing SSB.
However, for a nonrelativistic system it becomes much more complicated to clarify the connection between $N_{BG}$ and the number of GMs, as reflected in early work by Nielsen and Chadha~\cite{nielsen}.
Since then, much effort has been made to establish a proper classification of GMs~\cite{schafer, miransky, nicolis,nicolis2,  brauner-watanabe, watanabe, watanabe2, NG,NG2,NG3}.
A relevant point is the early debate between Anderson~\cite{anderson} and Peierls~\cite{peierls1, peierls2} regarding whether or not a ferromagnetic state should be considered as a result of SSB, now understood to be a paradigmatic example of SSB.
An insightful observation made by Nambu~\cite{nambu} led to the classification of type-A and type-B GMs~\cite{watanabe,watanabe2,NG,NG2,NG3} by introducing the Watanabe-Brauner matrix~\cite{brauner-watanabe}.
As a result, when the symmetry group $G$ is spontaneously broken into $H$, the counting rule for the GMs is established as
 \begin{equation}
N_A+2N_B=N_{BG}, \label{nab}
 \end{equation}
where $N_A$ and $N_B$ are the numbers of type-A and type-B GMs, and $N_{BG}$ is equal to the dimension of the coset space $G/H$.

A recent further development in another context concerns a connection between the entanglement entropy and the counting rule (\ref{nab}) for both type-A~\cite{Metlitski,Rademaker} and type-B~\cite{shiqianqian,dyw} GMs.
The scaling behavior of entanglement entropy continues to receive considerable attention in seemingly unrelated fields, such as black hole physics, quantum information science and quantum many-body physics. 
For quantum-many body systems, a key quantity is the entanglement entropy of ground states~\cite{ECP}. 
It is precisely in this context that a connection with GMs appears. 
In particular, the entanglement entropy $S(n)$ for a set of orthonormal basis  states, which appear as degenerate ground  states, scales logarithmically, in the thermodynamic limit, with the block size $n$ according to
 \begin{equation}
  S(n)=\frac{N_B}{2}\log_2n +S_{\!0}.
  \label{sf1f2}
  \end{equation}
The prefactor is precisely half the number of type-B GMs, thus leading to the identification of the number of type-B GMs with the fractal dimension $d_f$ exploited in Ref.~\cite{doyon} from
a field theoretic perspective, namely $d_f=N_B$, as far as the orthonormal basis states are concerned. As a consequence, previous findings about the entanglement entropy for the ${\rm SU}(2)$ ferromagnetic states are reproduced~\cite{popkov1,popkov2}.
This indicates that the highly degenerate ground states arising from SSB with type-B GMs are scale, but not conformally, invariant, reflecting the fact that there is an abstract fractal that lives in a Hilbert space. In a sense, such an interpretation for a ferromagnetic state as a scale invariant state challenges the folklore that scale invariance implies conformal invariance, as first put forward by Polyakov~\cite{polyakov}.

Insights towards understanding such unexpected connections can be gained from exactly solvable quantum many-body systems~\cite{baxterbook,giamarchi,sutherlandb,mccoy,sachdev}.
An example  is the spin-1 bilinear-biquadratic model~\cite{chubukov, fath1, fath2, fath3, kawashima, ivanov, rizzi, schmid, porras, romero, rakov, sierra1, sierra2, ronny, lundgren, dyw}, widely studied due to its prominent position in conceptual developments  in condensed matter physics.
A possible realization of the spin-1 bilinear-biquadratic model has been proposed via ultracold atoms in optical lattices~\cite{rodriguez, chiara,greiner1, jodens, Schneider}.
The model includes the Affleck-Kennedy-Lieb-Tasaki (AKLT) point~\cite{aklt1, aklt2}, which is a key to unlocking the mysteries behind the Haldane phase in the vicinity of the spin-1 antiferromagnetic Heisenberg model -- a prototypical example for symmetry-protected topological phases~\cite{spt1, spt2, spt3, spt4}.
The general model also includes other exactly solvable points: the $\rm{SU}(3)$ ferromagnetic and antiferromagnetic models~\cite{uimin, lai,sutherland2}, the spin-1 Takhtajan-Babujian model~\cite{tb1,tb2} and the ferromagnetic and antiferromagnetic spin-1 biquadratic models~\cite{barber,klumper}.
Among them the latter models are also of interest because they constitute a representation of the Temperley-Lieb algebra~\cite{tla,martin,baxterbook}, which in turn is relevant to the Jones polynomial in knot theory~\cite{jone}.
Each of the spin-1 Uimin-Lai-Sutherland, Takhtajan-Babujian and biquadratic models fit within the powerful framework of Yang-Baxter integrability in mathematical physics~\cite{baxterbook,sutherlandb,mccoy}.
Moreover, a spin-1 biquadratic model is mapped to a nine-state Potts model~\cite{barber}, but the underlying physics is quite different.
They share the same energy spectrum, but not level degeneracies, due to different symmetries.

For the $\rm{SU(3)}$ ferromagnetic spin-1 model~\cite{uimin, lai,sutherland2}, there is a sequence of highly degenerate and highly entangled ground states~\cite{popkov1,popkov2}, as a result of SSB from $\rm{SU(3)}$ to $\rm{SU(2)}\times\rm{U(1)}$, with two type-B GMs~\cite{shiqianqian}, if a ground state, over which the expectation values of the local order parameters are taken, is restricted to the highest weight state.
Thus the SSB pattern from $\rm{SU(3)}$ to $\rm{SU(2)}\times\rm{U(1)}$ occurs in the $\rm{SU(3)}$ ferromagnetic spin-$1$ model~\cite{shiqianqian}, resulting in four broken generators, accompanied with two type-B GMs.
Hence the counting rule (\ref{nab}) is satisfied, with $N_A=0$, $N_B=2$ and $N_{BG}=4$.

As noted above, both the uniform $\rm{SU(3)}$ ferromagnetic spin-1 model and the ferromagnetic spin-1 biquadratic model are two exactly solvable points of the more general spin-1 bilinear-biquadratic model.  An
intriguing observation is that the spin-1 biquadratic model is subject to an odd-even parity effect, as far as the symmetry group is concerned. Indeed, the symmetry group is the staggered $\rm{SU(3)}$ group if the system size is even and is the uniform  $\rm{SU(2)}$ group if the system size is odd, when periodic boundary conditions (PBCs) are adopted. In contrast, the symmetry group is always the staggered $\rm{SU(3)}$ group, when open boundary conditions (OBCs) are adopted.
Given they share $\rm{SU(2)}$ ferromagnetic states as degenerate ground states, one may anticipate that SSB with type-B GMs also emerges in the ferromagnetic spin-1 biquadratic model.
Note that the aforementioned even-odd parity effect implies that the SSB pattern is from ${\rm SU}(3)$ to ${\rm U}(1)\times {\rm U}(1)$ with two type-B GMs if the system size $L$ is even or from  ${\rm SU}(2)$ to ${\rm U}(1)$ with one type-B GM if the system size $L$ is odd, when PBCs are adopted. That is, we find that $d_f = N_B$, with $N_B = 2$, if $L$ is even and  $N_B = 1$, if $L$ is odd.
In fact, we are able to construct sequences of degenerate ground states that admit an exact Schmidt decomposition, leading to the unveiling of the underlying self-similarities in real space and implying the existence of an abstract intrinsic fractal~\cite{cantorset} characterized in terms of the fractal dimension $d_f$~\cite{doyon}.
A scaling analysis of the entanglement entropy (\ref{sf1f2}) makes it possible to identify the fractal dimension $d_f$ with the number of the type-B GMs $N_B$, if one {\it only} restricts to the orthonormal basis states constructed from the action of the lowering operator(s) on the highest weight state. For brevity, we mainly focus on the staggered $\rm{SU(3)}$ ferromagnetic spin-$1$ biquadratic model when $L$ is even if PBCs are adopted, unless otherwise stated.

For the staggered $\rm{SU(3)}$ ferromagnetic spin-$1$ biquadratic model, six of the eight generators are spontaneously broken, implying that the two type-B GMs originate from four broken generators,  with the two other generators being redundant, as required to keep consistency with the counting rule.
That is, there is an apparent difference between the two SSB patterns, from  $\rm{SU(3)}$ to $\rm{SU(2)}\times\rm{U(1)}$ and from $\rm{SU(3)}$ to  $\rm{U(1)}\times\rm{U(1)}$.
The former is an illustrative example for the counting rule~\cite{watanabe}, with four broken generators yielding two type-B GMs.
In contrast, for the latter, six generators are spontaneously broken, thus apparently violating the counting rule.
This justifies the necessity to introduce a redundancy principle for type-B GMs,  similar to SSB with type-A GMs~\cite{typeared}.

The finite-size ground states for the ferromagnetic spin-1 biquadratic model are highly degenerate, with the degeneracies being exponential with the system size.
As a consequence, the residual entropy per site is non-zero~\cite{spins}.
We find that the ground state degeneracies constitute the two Fibonacci-Lucas sequences, depending on either OBCs or PBCs.
As a result the ground state degeneracies are asymptotically the golden spiral -- an extremely well known self-similar geometric object.

With the above findings in view, we  turn now to the details of our calculations. In Section~\ref{su3spin1} we define the ferromagnetic spin-1 biquadratic model. Then in Section~\ref{Sec3} we discuss the underlying SSB and relevant GMs. This is followed in Section~\ref{Sec4} by exploration of the ground state degeneracies and their counting in terms of the Fibonacci-Lucas sequences, resulting in a nonzero residual entropy per site. The construction of a set of orthonormal basis states of an irreducible representation space of the symmetry group, which appear as degenerate ground states, is discussed in Section~\ref{Sec5}. The Schmidt decomposition of these states and other atypical degenerate ground states necessary for the calculation of their entanglement entropy and scaling behavior is given in Section~\ref{Sec6}. 
The identification of the fractal dimension with the number of GMs for sequences of degenerate ground states from the entanglement entropy scaling then follows in Section~\ref{Sec7}. A summary and concluding remarks follow in Section~\ref{Sec8}. Practicable technical considerations and finer details are relegated to the Appendices and Supplementary Material.

\section{The ferromagnetic spin-1 biquadratic model}
\label{su3spin1}

The ferromagnetic spin-1 biquadratic model is described by the Hamiltonian
\begin{equation}
H=\sum_{j} (\textbf{S}_j \cdot \textbf{S}_{j+1})^2. \label{hambq}
\end{equation}
Here $\textbf{S}_j=(S^x_{j},S^y_{j},S^z_{j})$ 
is the spin-1 operator at site $j$, with 
\begin{equation}
    S^x = \frac{1}{\sqrt 2}
    \begin{pmatrix} 0 & 1 & 0 \\ 
    1 & 0 & 1 \\
    0 & 1 & 0 
    \end{pmatrix}, \quad 
    S^y = \frac{1}{\sqrt 2} 
    \begin{pmatrix} 0 & -\mathrm{i} & 0 \\ 
    \mathrm{i} & 0 & -\mathrm{i} \\
    0 & \mathrm{i} & 0 
    \end{pmatrix}, \quad  
    S^z =  \begin{pmatrix} 1 & 0 & 0 \\ 
    0 & 0 & 0 \\
    0 & 0 & -1 
    \end{pmatrix} \,.
\end{equation}
As usual the sum over $j$ is taken from 1 to $L-1$ for OBCs, and from 1 to $L$ for PBCs.
Unless otherwise stated, we assume that the system size $L$ is even.
Generically, the model  (\ref{hambq}) possesses the staggered symmetry group ${\rm SU}(3)$, with its generators the
eight traceless matrices $J_\eta = \sum_j J_{\eta,j}$, for $\eta = 1, \ldots, 8$, as given in Appendix~\ref{AppA}.

The exact solvability~\cite{barber,klumper} of the spin-1 biquadratic model originates from the Hamiltonian (\ref{hambq}) constituting (up to an additive constant) a spin-1 representation of the Temperley-Lieb algebra~\cite{tla,baxterbook,martin}, and follows from a solution to the Yang-Baxter equation~\cite{baxterbook,sutherlandb,mccoy}.
As we shall see in detail below, 
a peculiar feature of the ferromagnetic spin-1 biquadratic model is that the ground states are highly degenerate, exponential with the system size $L$, thus leading to non-zero residual entropy per site~\cite{spins}.  Such highly degenerate ground states also occur for an anisotropic extension of the ferromagnetic spin-1 biquadratic model, for which the extended ground state phase diagram accommodates three symmetry-protected trivial phases, three coexisting fractal phases and six Luttinger liquid phases~\cite{cfJxyz}.

\section{SSB with type-B Goldstone modes and the redundancy principle} 
\label{Sec3}

The ferromagnetic spin-1 biquadratic model (\ref{hambq}) is subject to the odd-even parity effect if PBCs are adopted, because the symmetry group is the staggered $\rm{SU(3)}$ group if the system size $L$ is even and is the uniform  $\rm{SU(2)}$ group if the system size $L$ is odd, in contrast to the model under OBCs, with the symmetry group  being the staggered $\rm{SU(3)}$ group, irrespective of even or odd $L$. If  PBCs are adopted, then we have to distinguish odd and even $L$'s. When $L$ is odd, the symmetry group $\rm{SU(2)}$ is generated by $S^x$, $S^y$, and $S^z$, where 
$S^x=\sum_jS^x_j$, $S^y=\sum_jS^y_j$ and $S^z=\sum_j S^z_j$, with the SSB pattern being from $\rm{SU(2)}$ to $\rm{U(1)}$ generated by $S^z$, so the number of type-B GMs is one.
Accordingly, the coset space, denoted as $CP^{1}$, is diffeomorphic to the sphere $S^2$.

To investigate SSB and the appearance of GMs in the spin-1 biquadratic model when $L$ is even under PBCs, it is convenient to introduce the Cartan generators  $H_1$ and $H_2$, with the raising operators $E_1$, $E_2$, $E_3$ and the lowering operators $F_1$, $F_2$, $F_3$.
These generators and operators obey various commutation relations, as given in Appendix~\ref{AppA}, where they are also expressed in terms of the eight generators $J_\eta$. The same SSB pattern occurs under OBCs, irrespective of even or odd $L$.

If the highest weight state $|\rm{hws}\rangle$ is chosen to be $|\rm{hws}\rangle=|1...1\rangle$,  where  $|1\rangle$, together with  $|0\rangle$ and  $|\rm{-}1\rangle$, are the eigenvectors of $S^z_{j}$, with respective  eigenvalues $1$, $0$ and $-1$, 
 then the expectation values of the local components $H_{1,j}$ and  $H_{2,j}$ for the Cartan generators $H_1$ and $H_2$ are given by
$\langle1|H_{1,j}|1\rangle=1$ and $\langle1|H_{2,j}|1\rangle=1$, if $j$ is odd, and
$\langle1|H_{1,j}|1\rangle=0$ and $\langle1|H_{2,j}|1\rangle=1$, if $j$ is even.
It follows that for $E_\alpha$ and $F_\alpha$ ($\alpha=1$, $2$ and $3$), one may choose $F_{\alpha,j}$ and $E_{\alpha,j}$ as the interpolating fields~\cite{nielsen}, respectively.
Given $\langle[E_1,F_{1,j}]\rangle\propto \langle H_{1,j}\rangle$, $\langle[E_{1,j},F_1]\rangle\propto \langle H_{1,j}\rangle$,
$\langle[E_2,F_{2,j}]\rangle\propto \langle H_{2,j}\rangle$, $\langle[E_{2,j},F_2]\rangle\propto \langle H_{2,j}\rangle$,
$\langle[E_3,F_{3,j}]\rangle\propto \langle H_{1,j}-H_{2,j}\rangle$ and $\langle[E_{3,j},F_3]\rangle\propto \langle H_{1,j}-H_{2,j}\rangle$,
$E_\alpha$ and $F_\alpha$ are spontaneously broken, with $\langle H_{1,j}\rangle$ and $\langle H_{2,j}\rangle$ the local order parameters.
That is, six generators are spontaneously broken, but there are only two (linearly) independent local order parameters. This indicates that there are  two type-B GMs, and so $N_B=2$.
As a result of the Mermin-Wagner-Coleman theorem~\cite{mwc1,mwc2}, no type-A GM survives in one spatial dimension, thus the number of type-A GMs $N_A$ must be 0.
Hence, an apparent violation to the counting rule~(\ref{nab}) occurs for the SSB pattern from  $\rm{SU(3)}$ to  $\rm{U(1)}\times\rm{U(1)}$.
We remark that such a violation to the counting rule also occurs for SSB with type-A GMs~\cite{typeared}.
A well-known example is phonons in three spatial dimensions, which break both the translational symmetries and the rotational symmetries, resulting in six broken generators, with only three type-A GMs.

This justifies the necessity to introduce a redundancy principle, which has to be valid for both type-A and type-B GMs.
For our purpose, the redundancy principle may be stated as follows.
The number of GMs should not exceed the maximum number of (linearly) independent local order parameters.
As a result of the structure of Lie groups, this in turn implies that the number of type-B GMs should not exceed the maximum number of the commuting generators for the symmetry group, since an interpolating field is conjugate to the local component of its corresponding broken symmetry generator, which, together with a linear combination of the commuting generators, form a subgroup ${\rm SU(2)}$, as follows from the Watanabe-Brauner matrix~\cite{brauner-watanabe}.
This is consistent with a relevant mathematical theorem~\cite{schafer}. In particular, if the symmetry group is a semisimple Lie group, then the maximum number of the commuting generators becomes the rank $r$. In fact, if the symmetry group is not semisimple, then it must contain a semisimple subgroup that is  subject to SSB with type-B GMs.

In our case, the rank $r$ is equal to 2 for the staggered $\rm{SU(3)}$ symmetry group, then the number of type-B GMs must be equal to $N_B = 2$, implying that two broken generators are redundant for the SSB pattern from $\rm{SU(3)}$ to ${\rm U(1)}\times{\rm U(1)}$.

We stress that the presence of the odd-even parity effect in this model casts doubt on the existence of the thermodynamic limit itself. However, there is a way around this puzzle, if  type-B GMs are not {\it fundamental}, in the sense that they may be reduced to other low-lying excitations for both odd and even $L$'s. This will be addressed in a forthcoming article~\cite{greenpara}. Here we restrict ourselves to mention that the counting rule for the GMs, as presented in Eq.(\ref{nab}), is established under the assumption that any quantum many-body system undergoing SSB with the same symmetry group is described by the same effective low-energy theory~\cite{watanabe,watanabe2,NG,NG2,NG3}. This is a natural assumption that appears to work in a field-theoretic approach, which amounts to working directly in the thermodynamic limit.  However, our discussion above clearly shows that this is not the case (for finite sizes), since the symmetry group and the number of type-B GMs depend on what types of boundary conditions are adopted. Indeed, it is common to address SSB with type-B GMs in condensed matter systems when the system size $L$ is finite. In other words, the ferromagnetic spin-1 biquadratic model (\ref{hambq}) offers an example that illustrates the necessity to refine the counting rule.

\section{Fibonacci-Lucas sequences and the golden spiral}
\label{Sec4}

The SSB patterns from $\rm{SU(3)}$ to ${\rm U(1)}\times{\rm U(1)}$ or  from $\rm{SU(2)}$ to ${\rm U(1)}$ for the ferromagnetic spin-1 biquadratic model (\ref{hambq}) do not fully account for the exponential ground state degeneracies with $L$ if only the highest weight state is considered.
This is due to the fact that the dimension of the irreducible representation, generated from the highest weight state $|\rm{hws}\rangle$, for the staggered $\rm{SU}(3)$ group is  $(L+2)(3L+4)/8$ for $L$ even, and $(L+3)(3L+3)/8$ for $L$ odd  and $2L+1$ for the uniform $\rm{SU}(2)$ group.
Physically, not all generators of the staggered symmetry group $\rm{SU}(3)$ commute with an additional $Z_2$ symmetry operation, which is denoted as $\sigma$, the one-site translation operation $T_1$ under PBCs or the bond-centred inversion operation $I_b$ with $L$ even, and the site-centred inversion operation $I_s$ with $L$ odd, under OBCs.
Indeed, this is also reflected in the fact that the highest weight state $|1...1\rangle$ is the {\it only} one-site translation-invariant ground state under PBCs.

It is the presence of such an additional discrete $Z_2$ symmetry operation $\sigma$ that makes the ground state degeneracies exponential instead of being polynomial in $L$.
Indeed, the $Z_2$ symmetry operation $\sigma$ intertwines with the lowering operators $F_1$ and $F_2$, and they repeatedly act on the highest weight state $|1...1\rangle$ and the generalized highest weight states, thus yielding the entire ground state subspace (cf. Section I of the Supplementary Material).
This explains why the residual entropy per site is non-zero~\cite{spins}.
In other words, a generalized highest weight state becomes indispensable,  in addition to the highest weight state, to account for the exponential ground state degeneracies with $L$. Hence there should be a connection between additional discrete $Z_2$ symmetry operations and generalized highest weight states for such a quantum many-body system undergoing SSB with type-B GMs, if the ground state degeneracies under OBCs and PBCs are exponential.
Formally, the $m$-th generalized highest weight state $|{\rm hws}\rangle_g^m$  is defined recursively as $|{\rm hws}\rangle_g^m\notin V_0\oplus V_1\oplus... V_{m-1}$ and $E_1|{\rm hws}\rangle_g^m=0\mod(V_0\oplus V_1\oplus... V_{m-1})$, and $E_2|{\rm hws}\rangle_g^m=0 \mod(V_0\oplus V_1\oplus... V_{m-1})$, where $V_0$ denotes the subspace spanned by the degenerate ground states generated from the highest weight states, whereas $V_\mu$ denotes the subspace spanned by the degenerate ground states generated from the $\mu$-th generalized highest weight state, where $\mu=1,2,\ldots,m$.
Even though $|{\rm hws}\rangle_g^m$ are linearly independent to the states in the subspace  $V_0\oplus V_1\oplus \ldots V_{m-1}$, all the states $E_1^{M_1}E_2^{M_2}|{\rm hws}\rangle_g^m$ ($M_1\geq0$, $M_2\geq0$ and $M_1+M_2\geq1$) are linearly dependent to the states in the subspace $V_0\oplus V_1\oplus \ldots V_{m-1}$.

As an illustration, we show how to construct the degenerate ground states from the highest weight state and the generalized highest weight states for $L=2$, $4$ and $6$ in Section I of the Supplementary Material.
The necessity to introduce a generalized highest weight state lies in the fact that only the degenerate ground states generated from the highest weight state and the generalized highest weight states are amenable to further analysis,
which allows us to establish the scale invariance of the degenerate ground states, and to evaluate the entanglement entropy.

Here we briefly discuss how the presence of the generalized highest weight states reconciles with the odd-even parity effect under PBCs.
This is seen from the observation that, instead of being staggered ${\rm SU(3)}$ for even $L$, the symmetry group reduces to uniform ${\rm SU(2)}$, generated from $S^x$, $S^y$ and $S^z$, for odd $L$, if PBCs are adopted.
As a consequence, there is only one type-B GM, arising from the SSB pattern, namely ${\rm SU(2)}\rightarrow {\rm U(1)}$,  if $L$ is odd under PBCs. This is in sharp contrast to two type-B GMs arising from the SSB pattern ${\rm SU(3)}\rightarrow {\rm U(1)}\times {\rm U(1)}$ if $L$ is even.
However, in both cases, the generalized highest weight states are always present, with their total number being exponential, so that the ground state degeneracies are exponential with $L$, regardless of $L$ being even or odd.

The exponential ground state degeneracies with $L$ are also relevant to the fact that a fractal structure underlies the highly degenerate ground states, as follows from the possibility for an exact Schmidt decomposition (see below)~\cite{shiqianqian}.
Given that a wave function itself is not observable, it is necessary to choose a physical observable to reveal the scale invariance underlying such a fractal.
Indeed, if the residual entropy per site $S_{\!r}$ is chosen as an observable, then the underlying self-similar geometric object turns out to be the golden spiral.
That is, the ground state degeneracies ${\rm dim }(\Omega_L^{\rm OBC})$ and ${\rm dim }(\Omega_L^{\rm PBC})$  must be exponential with $L$, when $L\gg1$, with $\Omega_L^{\rm OBC}$ and $\Omega_L^{\rm PBC}$  being the ground state subspaces under OBCs and under PBCs.
Here we remark that an exponential function with an irrational base is involved, which must be less than $3$, but greater than $2$, since a simple estimate indicates that the ground state degeneracies are less than $3^{L}$, but greater than $2^{L}$.
Given  ${\rm dim }(\Omega_L^{\rm OBC})$ and ${\rm dim }(\Omega_L^{\rm PBC})$ are integer valued, we encounter a situation to express an integer in terms of an irrational number -- the so-called Binet formula~\cite{binet}.
A plausible surmise is that the ground state degeneracies satisfy a three-term recursive relation ${\rm dim }(\Omega_{L+2})=P \, {\rm dim }(\Omega_{L+1})-Q\,{\rm dim }(\Omega_{L})$  for $\Omega_L=\Omega_L^{\rm OBC}$ or $\Omega_L^{\rm PBC}$, with a characteristic equation being the quadratic equation $X^2-PX+Q=0$, where $P$ and $Q$ are co-prime integers: $(P,Q)=1$.
In fact, we are led to $P=3$ and $Q=1$ (for the details, see Appendix~\ref{AppB}).
As a consequence, we have
\begin{align}
{\rm dim }(\Omega_L^{\rm OBC})&=(R^{-2L-2}-R^{2L+2})/(R^{-2}-R^{2}), \quad L\geq2 \cr
{\rm dim }(\Omega_L^{\rm PBC})&=R^{-2L}+R^{2L}, \quad  L\geq3.
\end{align}
Here $R=(\sqrt{5}-1)/2$ is the (inverse) golden ratio.
More precisely, the ground state degeneracies  ${\rm dim }(\Omega_L^{\rm OBC})$ and  ${\rm dim }(\Omega_L^{\rm PBC})$ constitute two Fibonacci-Lucas sequences.
Hence the residual entropy per site is $S_{\!r}=-2 \log R$.
This is consistent with the previous results~\cite{spins}.
In fact, the three-term recursive relations for the two Fibonacci-Lucas sequences are identical to those for the ground state degeneracies~\cite{spins,saleur,moudgalya,katsura} when $L\geq2$ for OBCs and $L\geq 3$ for PBCs.
Physically, the non-zero residual entropy per site measures the disorder arising from the intertwining nature of the combined action of the additional $Z_2$ symmetry operation $\sigma$, and the lowering operators $F_1$ and $F_2$ on the highest weight state as well as the generalized highest weight states (cf. Section I of the Supplementary Material).

\section{Degenerate ground states from the highest weight state and the generalized highest weight states:  a hierarchical structure}
\label{Sec5}

For the ferromagnetic spin-1  biquadratic model under consideration, sequences of the highly degenerate ground  states are generated from the repeated action of $F_1$ and $F_2$, combining with the additional $Z_2$ symmetry operation $\sigma$ and the time-reversal operation $K$, on both the highest weight state and the generalized highest weight states.  Here we {\it only} focus on the Hamiltonian (\ref{hambq}) when the symmetry group is the staggered ${\rm SU(3)}$ group, as it happens if OBCs are adopted and if $L$ is even when PBCs are adopted. The construction of sequences of the highly degenerate ground  states is simpler if $L$ is odd when PBCs are adopted, since they are generated from the repeated action of $S^-$ on the highest weight state and the generalized highest weight states.

As it turns out, a hierarchical structure arises among exponentially many degenerate ground states. Indeed, this hierarchical structure has already appeared in exponentially many generalized highest weight states,  including the highest weight state as a special case.  For our purpose, one may define  {\it atypical} (periodic) generalized highest weight states as degenerate factorized  ground states, if they consist of local states $ \vert 1_{j}\rangle$ and  $ \vert \,0_{j}\rangle$, in the sense that a {\it typical} degenerate factorized ground state is non-periodic, even if local states $\vert - \!1_{j}\rangle$ are excluded. With this definition in mind, atypical degenerate  ground states are defined to be those degenerate ground states generated from the action of the lowering operators on atypical generalized highest weight states.
Physically, the occurrence of atypical degenerate ground  states with period $p$ may be understood as follows. SSB occurs from the symmetry under the one-site translation operation to a
discrete symmetry group $\bigoplus_p \mathscr{Z}_p$, with $\mathscr{Z}_p$ generated by the $p$-site translation operation under PBCs, which always accompanies SSB from  ${\rm SU(3)}$ to ${\rm U(1)} \times {\rm U(1)}$. Here the sum  $\bigoplus_p$ is taken over all possible $p$'s if $p$ divides $L$.
Note that the number of atypical degenerate ground states depends on $L$, since $p$ depends on $L$. As such, the exponential ground state degeneracies with $L$ relate to this dependence of $p$ on $L$.

Generically,  an atypical generalized highest weight state, with the period   $p$ being an integer, may be written as\\ $|s_1s_2\ldots s_p \ldots s_1s_2 \ldots s_p\rangle$, where $s_1,s_2, \ldots, s_p$ are either $|1\rangle$ or $|0\rangle$, but subject to the condition that no two adjacent sites $j$ and $j+1$ are occupied by the local states $|0\rangle_j$ and $|0\rangle_{j+1}$.
Then, a sequence of the degenerate ground states, with the system size $L$ a multiple of $p$, are generated from the repeated action of  two of the three lowering operators $F_1$, $F_2$ and $F_3$ on such a generalized highest weight state $|s_1s_2\ldots s_p \ldots s_1s_2 \ldots s_p\rangle$, given $F_3=T_1F_1T_1$ under PBCs and $F_3=I_bF_1I_b$ under OBCs.
We remark that the highest weight state itself $|1...1\rangle$ is exceptional, in the sense that it may be regarded as a special case, with the period $p=1$.
As a result, the degenerate ground states generated from the highest weight states have periodicity $2$, due to the staggered structure of the lowering operators $F_1$, $F_2$ and $F_3$.

We may choose the two lowering operators $F_1$ and $F_2$ as an example.
A sequence of atypical degenerate ground states, denoted as  $|L,M_1,M_2\rangle_p$, are hence generated from the repeated action of the two lowering operators $F_1$ and $F_2$ on an atypical generalized highest weight state $|s_1s_2\ldots s_p \ldots s_1s_2 \ldots s_p\rangle$:
\begin{equation}
 |L,M_1,M_2\rangle_p=\frac{1}{Z_p(L,M_1,M_2)}F_1^{M_1}F_2^{M_2}|s_1s_2\ldots s_p \ldots s_1s_2 \ldots s_p\rangle,
 \label{lm1m2p}
\end{equation}
where $Z_p(L,M_1,M_2)$ is introduced to ensure that  $|L,M_1,M_2\rangle_p$ is normalized.
We emphasize that $M_1$ and $M_2$ are subject to some constraints, which may be related to  $L$.

Here we also emphasize that an atypical generalized highest weight state with the period $p$ is invariant under the permutation group $S_{L/p}$, if $p$ divides $L$. 
In particular, the highest weight state is invariant under the permutation group $S_{L}$, if it is chosen properly to ensure that $p=1$. 
In contrast, a {\it typical} degenerate ground state is not permutation-invariant, thus pointing towards a connection to Green parafermions that remains elusive up to the present since they were introduced decades ago~\cite{green}, as far as their realizations in condensed matter systems are concerned, given that Green parafermions constitute high-dimensional representations of the permutation group $S_{L}$.  We shall return to this intriguing topic in a forthcoming article~\cite{greenpara}.  Here we stress that not all generalized highest weight states are atypical, as defined for the model under investigation. However, there is an emergent subsystem non-invertible symmetry that acts  on all atypical generalized highest weight states to yield all highest weight states, as demonstrated in Ref.~\cite{greenpara}.  In other words, there is a hierarchical structure behind all generalized highest weight states according to their behavior under permutations, which in turn leads to a hierarchical structure behind  exponentially many degenerate ground states.

In the thermodynamic limit when $L$ tends to infinity, it is plausible to take $p=1, 2, \ldots, \infty$. Hence, we are able to generate {\it atypical}  degenerate ground states by acting lowering operator(s) on atypical generalized highest weight states. A notable result is that among an ensemble of highly degenerate atypical (periodic) ground states in the thermodynamic limit, there are a monomerized ground state with $p=1$, dimerized ground states with $p=2$, trimerized ground states with $p=3$, tetramerized ground states with $p=4$, and so on. This  is in sharp contrast to the antiferromagnetic spin-1 biquadratic model, where only dimerized ground states are involved~\cite{afflecksun1}, thus offering a way to characterize this novel type of quantum state of matter.

Given that it is a formidable task to evaluate the norms for exponentially many  degenerate ground states, we {\it only} focus on atypical degenerate ground states, which fall into two classes of degenerate ground states: one is degenerate ground states generated from the highest weight state, and the other is degenerate ground states generated from special periodic generalized highest weight states. Specifically, we restrict ourselves to two atypical (periodic) generalized highest weight states, with periods $p=4$ and $p=6$, in addition to the highest weight state.

\subsection{Degenerate ground states generated from the highest weight state}

There are distinct choices for the highest weight state $|\rm{hws}\rangle$, two of which are as follows.
One is the translation-invariant factorized ground state: $|\rm{hws}\rangle=|1...1\rangle$, the other is the factorized ground state without the one-site translational invariance: $|\rm{hws}\rangle=|0_x0_y...0_x0_y\rangle$, under PBCs.
Here $|0_x\rangle$ and $|0_y\rangle$ are the eigenvectors of $S^x_j$ and $S^y_j$, with respective eigenvalues $0$.
We remark that the two choices are unitarily equivalent under a local unitary transformation $U$, satisfying $K_x=UK_yU^{\dagger}$,  $K_y=UK_zU^{\dagger}$ and  $K_z=UK_xU^{\dagger}$, with $K_x=J_5$, $K_y=J_6$ and $K_z=J_3$ being the three generators of a symmetry subgroup ${\rm SU}(2)$.
Here $U$ takes a product form $\prod_jU_j$, where 
$U_j$ is given in (\ref{eqnsA3}).

Note that other choices may be induced from the symmetric group ${\rm S}_3$, arising from the cyclic permutations with respect to $x$, $y$, and $z$.
As a consequence, the entanglement entropy  $S_f(L,n)$ (cf. the next Section  for its definition) for distinct sequences of the degenerate ground states, corresponding to the different choices for the highest weight state $|\rm{hws}\rangle$, must be identical.
Therefore, we only need to focus on one choice: $|\rm{hws}\rangle=|1...1\rangle$.

A sequence of the degenerate ground states $|L,M_1,M_2\rangle_2$ ($M_1=0$, \ldots, $L/2$ and $M_2=0$, \ldots, $L$) in a two-site periodic structure, are generated from the repeated action of the lowering operators $F_1$ and $F_2$ on the highest weight state $|1...1\rangle$:
 \begin{equation}
 |L,M_1,M_2\rangle_2=\frac{1}{Z_2(L,M_1,M_2)}F_1^{M_1}F_2^{M_2}|1...1\rangle,
 \label{lm1m2}
 \end{equation}
where  $Z_2(L,M_1,M_2)$ is introduced to ensure that $|L,M_1,M_2\rangle_2$ is normalized. It takes the form
\begin{equation}
Z_2(L,M_1,M_2)=M_1!M_2!\sqrt{C_{L/2}^{M_1}C_{L-M_1}^{M_2}}.\label{zm1m2}
\end{equation}
A derivation of the concrete expression for $Z_2(L,M_1,M_2)$ is given in Section II of the Supplementary Material.

In fact, in addition to $|L,M_1,M_2\rangle_2$ in Eq.~(\ref{lm1m2}),  a sequence of the degenerate ground states $|L,M_1,M_3\rangle_2$ ($M_1=0$, \ldots, $L/2+M_3$ and $M_3=0$, \ldots, $L/2$), are generated from the repeated action of the lowering operators $F_1$ and $F_3$ on the highest weight state $|1...1\rangle$:
\begin{equation}
 |L,M_1,M_3\rangle_2=\frac{1}{Z_2(L,M_1,M_3)}F_1^{M_1}F_3^{M_3}|1...1\rangle,
 \label{lm1m3}
 \end{equation}
where  $Z_2(L,M_1,M_3)$ is introduced to ensure that $|L,M_1,M_3\rangle_2$ is normalized. It takes the form
\begin{equation}
Z_2(L,M_1,M_3)=M_1!M_3!\sqrt{C_{L/2}^{M_3}C_{L/2+M_3}^{M_1}}.
\label{zm1m3}
\end{equation}
A derivation of this concrete expression for $Z_2(L,M_1,M_3)$ is given in Section II of the Supplementary Material.
The degenerate ground states, generated from the highest weight state $|\rm{hws}\rangle=|0_x0_y...0_x0_y\rangle$ are discussed in Section III of the Supplementary Material.

The emergence of a bunch of the generalized highest weight states is not only responsible for the exponential ground state degeneracies with $L$, but also accounts for the difference between the ground state degeneracies under OBCs and PBCs (details are given in Section I of the Supplementary Material).

For our purpose, we choose two typical generalized highest weight states: $|1110...1110\rangle$ and $|10\rm{-}1\rm{-}101...10\rm{-}1\rm{-}101\rangle$, which exhibit a four-site and six-site periodic structure, i.e., with $p=4$ and $p=6$.

A sequence of the degenerate ground states $|L,M_2,M_3\rangle_4$ ($M_2=0, \ldots , 3L/4$ and $M_3=0, \ldots, L/4$), in a four-site periodic structure, are generated from the repeated action of the lowering operators $F_2$ and $F_3$ on $|1110...1110\rangle$, with period $p=4$:
 \begin{equation}
 |L,M_2,M_3\rangle_4=\frac{1}{Z_4(L,M_2,M_3)}F_2^{M_2}F_3^{M_3}|1110...1110\rangle,
 \label{lm2m3p4}
 \end{equation}
 where $Z_4(L,M_2,M_3)$ is introduced to ensure that  $|L,M_2,M_3\rangle_4$ is normalized. It takes the form
\begin{equation}
Z_4(L,M_2,M_3)=M_2!M_3!\sqrt{C_{L/4}^{M_3}C_{3L/4-M_3}^{M_2}}.
\label{zm2m3p4}
\end{equation}
A derivation of $Z_4(L,M_2,M_3)$  is given in Section II of the Supplementary Material. 

In the six-site periodic structure, a sequence of the degenerate ground states $|L,M_1,M_2\rangle_6$ ($M_1=0$, \ldots , $L/3$ and $M_2=0$, \ldots , $L/3$) 
are generated from the repeated action of the lowering operators $F_1$ and $F_2$ on
 $|10\rm{-}1\rm{-}101...10\rm{-}1\rm{-}101\rangle$, with period $p=6$:
\begin{equation}
 |L,M_1,M_2\rangle_6=\frac{1}{Z_6(L,M_1,M_2)} 
 F_1^{M_1}F_2^{M_2}|10\rm{-}1\rm{-}101...10\rm{-}1\rm{-}101\rangle,
 \label{lm1m2p6}
\end{equation}
where  $Z_6(L,M_1,M_2)$ is introduced to ensure that  $|L,M_1,M_2\rangle_6$ is normalized. It takes the form
\begin{equation}
Z_6(L,M_1,M_2)=M_1!M_2!\sqrt{\sum_{l=0}^{M_1}C_{L/6}^{l}C_{L/6}^{M_1-l}C_{L/3-l}^{M_2}} \,.
\label{zm1m2p6}
\end{equation}
A derivation of $Z_6(L,M_1,M_2)$ is given in Section II of the Supplementary Material.

Similarly, a Schmidt decomposition may be performed for the degenerate ground states $|L,M_1,M_3\rangle_2$,
$|L,M_2,M_3\rangle_4$ and $|L,M_1,M_2\rangle_6$ given in equations (\ref{lm1m3}), (\ref{lm2m3p4}) and (\ref{lm1m2p6}) (as discussed in Section IV of the Supplementary Material).

\section{Schmidt decomposition for atypical degenerate ground states}
\label{Sec6}

Both atypical and  typical degenerate ground states share a common feature that an exact Schmidt decomposition is admitted, which may be exploited to facilitate the evaluation of  the entanglement entropy. Indeed, this reveals a hierarchical structure behind exponentially many (linearly independent) degenerate ground states, which are generated from the action of lowering operators $F_1$, $F_2$, and $F_3$, combining with the additional  $Z_2$ symmetry operation $\sigma$ and the time-reversal operation $K$, on the highest weight state as well as the generalized highest weight states, if the symmetry group is the staggered ${\rm SU(3)}$ group. 
The core is a subspace within the ground state subspace, which is spanned by polynomially many  degenerate ground states generated from the highest weight state. Mathematically, this subspace is an irreducible representation space for the symmetry group. In fact, these states constitute orthonormal basis states in this particular subspace, in contrast to other degenerate ground states generated from  generalized highest weight states.

We are now in a position to show that such a generic atypical degenerate ground state admits an exact Schmidt decomposition, if the system is bipartitioned  into a block of size $n$ and its environment of size $L-n$, thus implying the self-similarities underlying the degenerate ground states as a whole.
Here we restrict ourselves to two atypical degenerate ground states, with the period $p$ being four and six, respectively, in addition to the highest weight state.

The atypical degenerate ground states $|L,M_1,M_2\rangle_p$, with $p$ even,  in Eq.~(\ref{lm1m2p}) admit an exact Schmidt decomposition, with
\begin{equation}
|L,M_1,M_2\rangle_p= \sum\limits_{k_1}\sum\limits_{k_2}\lambda(L,n,k_1,k_2,M_1,M_2)
|n,k_1,k_2\rangle_p|L-n,M_1-k_1,M_2-k_2\rangle_p, \label{svd}
\end{equation}
where $n$ and $L$ are multiple of $p$, and the Schmidt coefficients $\lambda(L,n,k_1,k_2,M_1,M_2)$ take the form
\begin{equation}
\lambda(L,n,k_1,k_2,M_1,M_2)=C_{M_1}^{k_1}C_{M_2}^{k_2}\frac{Z_p(n,k_1,k_2)Z_p(L-n,M_1-k_1,M_2-k_2)}{Z_p(L,M_1,M_2)}.
\label{m1m2p}
\end{equation}
Here $Z_p(n,k_1,k_2)$ and $Z_p(L-n,M_1-k_1,M_2-k_2)$ take the same form as $Z_p(L,M_1,M_2)$.

Before proceeding, we briefly discuss an exact Schmidt decomposition for typical (non-periodic) degenerate ground state generated from the action of the lowering operators on a typical (non-periodic) generalized highest weight state $|s_1s_2\ldots s_L\rangle$. Here we have assumed that $s_j$ ($j=1,2,\ldots,L$) are $1$ or 0, subject to the constraint that there are no two adjacent sites $j$ and $j+1$, on which $s_j$ and $s_{j+1}$ are 0 so that it is a factorized ground state under PBCs or OBCs. A simple calculation shows that an exact Schmidt decomposition is allowed for a typical (non-periodic) degenerate ground state, thus indicating the self-similarities in real space underlying the entire ground state subspace, as argued in Refs.~\cite{shiqianqian,cantorset}.
However, as already mentioned above, typical (non-periodic) degenerate ground states are not invariant under permutations, so their scaling behavior of the entanglement entropy is much more complicated. Fortunately, our main concern is about the  scaling behavior of the entanglement entropy in the thermodynamic limit when the block size $n$ is large enough and the system size $L$ tends to infinity. In this case, it is plausible to demand that all atypical generate ground states, with the period $p$ being any integer, are allowed. In particular, as shown in Appendix~\ref{AppC}, the set for all the fillings of atypical generalized highest weight states is dense with respect to all possible fillings in the thermodynamic limit. In this sense, it is sufficient to focus on the atypical degenerate ground states generated from the action of the lowering operators on atypical generalized highest weight states.

\section{The entanglement entropy: identification of the fractal dimension with the number of GMs for orthonormal basis states and beyond}
\label{Sec7}

The usefulness of an exact Schmidt decomposition lies in the fact that 
for any atypical degenerate ground state $|L,M_1,M_2\rangle_p$ with the period $p$, the entanglement entropy $S_{\!L}(n,M_1,M_2)$ follows from
\begin{align}
S_L(n,M_1,M_2)=-\sum_{k_1,k_2} \Lambda(L,n,k_1,k_2,M_1,M_2)
\log_{2}\Lambda(L,n,k_1,k_2,M_1,M_2),
\label{snkm1m2}
\end{align}
where $\Lambda(L,n,k_1,k_2,M_1,M_2)$ are the eigenvalues of the reduced density matrix $\rho_L(n,M_1,M_2)$, given by\\
$\Lambda(L,n,k_1,k_2,M_1,M_2)=[\lambda(L,n,k_1,k_2,M_1,M_2)]^2$.
This greatly simplifies the evaluation of the entanglement entropy, thus in turn making it possible to perform a systematic analysis of the scaling behavior of the entanglement entropy for both the orthonormal basis states  in the (irreducible representation) space of the symmetry group and beyond. Note that  the orthonormal basis states  are generated from the highest weight state~\cite{{shiqianqian},finitesize}, so the period $p$ is either one or two. In addition, we shall extend our discussion of this scaling behavior to other atypical (periodic) degenerate ground states, with their periods being greater than two, subject to the condition that both the block size $n$ and the system size $L$ are multiple of the period $p$. Given that an atypical degenerate ground state $|L,M_1,M_2\rangle_p$ is invariant under the permutation group $S_{L/p}$, so the bond- or site-centered inversion symmetry is not spontaneously broken for even or odd  $L$. As a result,  the entanglement entropy $S_{\!L}(n,M_1,M_2)$ satisfies the inversion symmetry relation
$S_{\!L}(n,M_1,M_2) = S_{\!L}(L-n,M_1,M_2)$, as long as both $n$ and $L$ are multiple of $p$.

To begin with, we stress that the argument in Ref.~\cite{finitesize} is valid for the orthonormal basis states in the irreducible representation space of the symmetry group within the ground state subspace, the entanglement entropy $S_{\!\!f}(L,n)$ scales in the form
\begin{equation}
S_{\!\!f}(L,n)=\frac{N_B}{2} \log_2\frac{n(L-n)}{L} +S_{\!\!f0}.
\label{slnf}
\end{equation}
Here  $S_{\!\!f0}$ is an additive non-universal constant  and the subscript
$f$ refers to a set of fillings $f_1, f_2$, and $f_3$ defined with respect to the highest weight state, or a set of relative fillings $f_1^*$, $f_2^*$ and $f_3^*$ defined with respect to an atypical generalized highest weight state.
We remark that the fillings $f_\alpha$  are assumed to be neither zero nor the maximum. In addition, one may expect that the scaling relation (\ref{slnf}) still works for all atypical (periodic) degenerate ground states, with the period $p$ being any integer that divides $L$, as follows from the same reasoning in Ref.~\cite{finitesize}.

Note that the finite-size scaling relation (\ref{slnf}) reduces to the scaling relation (\ref{sf1f2}) in the thermodynamic limit. More precisely, for fixed fillings $f_1=M_1/L$ and $f_2=M_2/L$ in the thermodynamic limit $L \rightarrow \infty$,  the entanglement entropy $S_{\!L}(n,M_1,M_2)$ becomes $S(n,f_1,f_2)$.
That is, the scaling of the entanglement entropy $S(n,f_1,f_2)$ with the block size $n$, in the thermodynamic limit, for any non-zero fillings $f_1$ and $f_2$, takes the same form as Eq.~(\ref{sf1f2}).

\begin{figure}[t!]
	\centering
\includegraphics[width=0.45\textwidth]{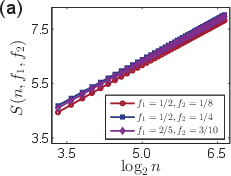}
\includegraphics[width=0.48\textwidth]{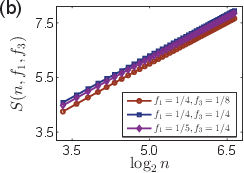}
\includegraphics[width=0.45\textwidth]{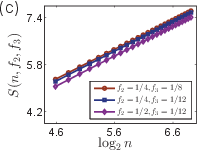}
\includegraphics[width=0.48\textwidth]{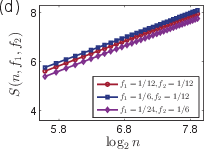}
	\caption{ (a) The entanglement entropy $S(n,f_1,f_2)$ vs $\log_2 n$ in the thermodynamic limit for different fillings
		   $f_1$ and $f_2$.
		The prefactor is close to $1$, within an error less than 1.2$\%$.
		(b) The entanglement entropy $S(n,f_1,f_3)$ vs $\log_2 n$ in the thermodynamic limit for different fillings $f_1$ and $f_3$.
		The prefactor is close to $1$, within an error less than 2$\%$.
		(c)  The entanglement entropy $S(n,f_2,f_3)$ vs $\log_2 n$ in the thermodynamic limit for different fillings $f_2$ and $f_3$. Here a generalized highest weight state is chosen to be $|1110...1110\rangle$.
		The prefactor is close to $1$, within an error less than 1.5$\%$.
		(d)  The entanglement entropy $S(n,f_1,f_2)$ vs $\log_2 n$ in the thermodynamic limit for different fillings $f_1$ and $f_2$. Here a generalized highest weight state is chosen to be $|10\rm{-}1\rm{-}101...10\rm{-}1\rm{-}101\rangle$.
		The prefactor is close to $1$, within an error less than 2$\%$.
	}
	\label{Su3many}
\end{figure}

Plots of the entanglement entropy against the block size in the thermodynamic limit are shown in Fig.~\ref{Su3many} for different filling factors.
The results imply that the contribution from the filling factors is a non-universal additive constant as in Eq.~(\ref{sf1f2}).
The results are thus consistent with the generic but heuristic argument given in Ref.~\cite{shiqianqian}. In addition, 
the emergence of the logarithmic scaling behavior of the entanglement entropy with the block size $n$, when the system size $L$ approaches the thermodynamic limit, is demonstrated in Appendix~\ref{AppD}.

We see that the entanglement entropy scales with block size $n$ for each of the three choices in the same way as Eq.~(\ref{sf1f2}), with $n$ being a multiple of two, a multiple of four and a multiple of six, respectively. Importantly, for each of the sets of highly degenerate ground states under consideration, the entanglement entropy is seen to scale with the block size, where the prefactor is the number of type-B GMs $N_B=2$.

We stress that this scaling behavior (\ref{slnf}) is a special case of a general scaling relation of the entanglement entropy for a linear combination of spin or generalized coherent states of the symmetry group on a support. As shown in Ref.~\cite{cantorset},  any support may be approximated by a fractal  decomposable into a set of the Cantor sets. Mathematically, we have
\begin{equation}
S_{\!\!f}(L,n)=\frac{d_f}{2} \log_2\frac{n(L-n)}{L} +S_{\!\!f0},
\label{slnf1}
\end{equation}
where $d_f$ denotes the fractal dimension for a support  to form a linear combination of  spin or generalized coherent states. As we have shown in Ref.~\cite{cantorset}, such a linear combination in the (irreducible representation) space of the symmetry group spanned by the above orthonormal basis  states {only} partially senses the presence of type-B GMs, since it may be re-interpreted as a linear combination of the orthonormal basis  states generated by the action of lowering operators on the highest weight state.

In Appendix~\ref{AppE}, we have numerically checked that the general scaling relation (\ref{slnf1}) works for a linear combination of spin or generalized coherent states, with support being a Cantor set, depending on the symmetry group due to the odd-even-parity effect. This confirmed our prediction in Ref.~\cite{cantorset}. In particular, the fractal dimension $d_f$ is identified with the number of type-B GMs $N_B$ for the orthonormal basis states in the irreducible representation space within the ground state subspace, valid for both odd and even $L$'s.
In addition, the same scaling relation (\ref{slnf}) has been confirmed for other atypical (periodic) degenerate ground states, with the periods being greater than two.

As for typical (non-periodic) degenerate ground states, it is straightforward to compute the entanglement entropy from an exact Schmidt decomposition. However, they are not invariant under permutations, thus meaning that the bond- or site-centered inversion symmetry is spontaneously broken for a generic typical (non-periodic) degenerate ground state, depending on even or odd $L$. As a consequence, the entanglement entropy $S_{\!L}(n,M_1,M_2)$ does not satisfy the inversion symmetry relation $S_{\!L}(n,M_1,M_2) = S_{\!L}(L-n,M_1,M_2)$. Hence their scaling behavior of the entanglement entropy is different from that for atypical degenerate ground states. We speculate that the difference is {\it only} manifested in a non-universal additive constant, which must depend on $n$ explicitly for typical (non-periodic) degenerate ground states. However, such a dependence on $n$ tends to be vanishing as $n$ gets large enough, in order to keep consistency with the fact that the set for all the fillings of atypical generalized highest weight states is dense in the entire filling set (cf. Appendix~\ref{AppC}).  In other words, the logarithmic scaling form remains the same, in the sense that, for typical degenerate ground states, only type-B GMs are sensed as low-lying excitations, though type-B GMs themselves are not fundamental.
We shall leave this subtle issue to a future publication.

\section{Summary and conclusion}
\label{Sec8}

We have shown that the physical and  mathematical structure underlying the highly degenerate ground states of the ferromagnetic spin-1 biquadratic model (\ref{hambq}) is particularly rich.
For this model the SSB pattern from $\rm{SU(3)}$ to $\rm{U(1)}\times\rm{U(1)}$ with two type-B GMs or from $\rm{SU(2)}$ to $\rm{U(1)}$ with one type-B GM, leads to scale, but not conformally, invariant degenerate ground states.
This indicates that  an abstract fractal emerges in the Hilbert space, with the fractal dimension being identical to the number of type-B GMs, as far as the orthonormal basis states in the irreducible representation space of the symmetry group and other atypical (periodic) degenerate ground states generated from the repeated action of the lowering operator(s) on the highest weight state and  the generalized highest weight states are concerned. As such, the presence of such a quantum state challenges the folklore that scale invariance implies conformal invariance~\cite{polyakov}, with its ramifications for our understanding of quantum critical phenomena remaining to be clarified~\cite{fm}.

We found it necessary to introduce a redundancy principle, to keep consistency with the counting rule~(\ref{nab}) for type-B GMs, as  also happens for type-A GMs.
Given the staggered nature of the symmetry group  ${\rm SU}(3)$ for even $L$ under PBCs or both even and odd $L$'s under OBCs, the ground state degeneracies are exponential with the system size $L$, which accounts for the fact that the residual entropy per site is  non-zero.
This may be explained based on an observation that the degenerate ground states exhibit a periodic structure,
which in turn  results from SSB of the one-site translation symmetry to the discrete symmetry group $\bigoplus_p \mathscr{Z}_p$ that always accompanies SSB of the staggered  ${\rm SU(3)}$ group or the uniform ${\rm SU(2)}$ group.
Indeed, we have seen that among the degenerate ground states of the ferromagnetic spin-1 biquadratic model there is monomerized ground states with period $p=1$, dimerized ground states with $p=2$, tetramerized ground states with $p=4$, and so on. 
This is in sharp contrast to the antiferromagnetic spin-1 biquadratic model, for which Affleck~\cite{afflecksun1} originally noted that only dimerized ground states are involved.

Remarkably, the ground state degeneracies for the spin-1 ferromagnetic biquadratic model constitute the two Fibonacci-Lucas sequences, thus providing an unexpected connection to an ancient mathematical gem.
As a consequence, the ground state degeneracies are asymptotically the golden spiral, a well known self-similar geometric object.
In addition, sequences of degenerate ground states generated from highest and generalized highest weight states have been constructed in Section~\ref{Sec5} to establish that the entanglement entropy scales logarithmically with the block size in the thermodynamic limit, with the prefactor being half the number of type-B GMs for atypical (periodic) degenerate ground states generated from highest and atypical generalized highest weight states.

{\it Note added.} There have been a number of related developments after this work was completed. In particular, some subtle issues have been clarified in Ref.~\cite{cantorset}, as far as the connection between the fractal dimension and the number of type-B GMs is concerned. In addition, a universal finite system-size scaling analysis of the entanglement entropy has been developed for highly degenerate ground states arising from SSB with type-B GMs in quantum many-body systems~\cite{finitesize}. The universal scaling form was demonstrated for the ${\rm SU}(2)$ spin-$s$ Heisenberg ferromagnetic model, the ${\rm SU}(N+1)$ ferromagnetic model, and the staggered ${\rm SU}(3)$ spin-1 ferromagnetic biquadratic model. In other work the appearance of type-B GMs and a logarithmic spiral was established for the staggered ${\rm SU}(4)$ ferromagnetic spin-orbital model~\cite{SU4}. The work reported here was also subsequently extended to the study of SSB, entanglement and logarithmic spirals in a quantum spin-1 many-body system with competing dimer and trimer interactions~\cite{DTmodel}. For the one-dimensional flat-band ferromagnetic Tasaki model entanglement entropy scaling reveals SSB with one type-B GM~\cite{Tasaki}. Universal finite-size scaling and entanglement entropy for a type of scale-invariant states in two spatial dimensions and beyond have also been investigated~\cite{2d}. In a further development it has been shown that the exponential ferromagnetic ground state degeneracies imply the emergence of Goldstone flat bands, and a connection with quantum many-body scars has been revealed~\cite{flatband}, which in turn are relevant to the projected Green parafermion states~\cite{greenpara}.

\section*{Acknowledgements}
We thank John Fjaerestad for enlightening discussions and
Sanjay Moudgalya and Hosho Katsura for bringing their work to our attention.

\section*{Appendices}
\onecolumngrid
\setcounter{section}{0}
\setcounter{equation}{0}
\setcounter{figure}{0}
\renewcommand{\theequation}{A\arabic{equation}}
\renewcommand{\thefigure}{A\arabic{figure}}
\subsection{Generators for the staggered symmetry group {SU(3)}}
\label{AppA}

In contrast to the uniform symmetry group ${\rm SU(3)}$ for the ${\rm SU(3)}$ ferromagnetic model, the generators of the staggered symmetry group ${\rm SU(3)}$ generically do not commute with the one-site translation operation under PBCs or the bond-centred inversion operation with $L$ even, and the site-centred inversion operation with $L$ odd under OBCs.

Different choices for the eight generators of the staggered symmetry group ${\rm SU(3)}$ may be made.
The eight generators mentioned in the main text: $J_\eta$, for $\eta= 1, \ldots, 8$, are given by~\cite{cfJxyz}
\begin{eqnarray*} 
& J_{1,j}=S^x_{j}/2, \quad J_{2,j}=S^y_{j}/2, \quad J_{3,j}=S^z_{j}/2 \\
&J_{4,j}=(-1)^j(1-3/2 \, (S^z_{j})^2), \quad J_{5,j}=(-1)^j((S^x_{j})^2-(S^y_{j})^2)/2, \\
&J_{6,j}=(-1)^j(S^x_{j} S^y_{j}+S^y_{j} S^x_{j})/2, \quad J_{7,j}=(-1)^j(S^y_{j} S^z_{j}+S^z_{j}S^y_{j})/2, \quad J_{8,j}=(-1)^j(S^z_{j} S^x_{j}+S^x_{j} S^z_{j})/2 \,. 
\end{eqnarray*}

In addition to the above generators, 
it is more convenient to work with the Cartan generators $H_1$ and $H_2$,  the raising operators $E_1$, $E_2$, $E_3$, and the lowering operators $F_1$, $F_2$, $F_3$.
We list two different choices for the staggered symmetry group ${\rm SU}(3)$ below.

(i) The first choice for the eight generators, commuting with the Hamiltonian (\ref{hambq}), is
\begin{eqnarray} \label{8gens}
&H_1=\sum_jH_{1,j}, \quad H_2=\sum_jH_{2,j}, \nonumber\\
&E_1=\sum_jE_{1,j}, \quad E_2=\sum_jE_{2,j}, \quad E_3=\sum_jE_{3,j},\\
&F_1=\sum_jF_{1,j}, \quad F_2=\sum_jF_{2,j}, \quad F_3=\sum_jF_{3,j}, \nonumber
\end{eqnarray}
where, for odd $j$,
\begin{eqnarray*}
&H_{1,j}= J_{3,j}+J_{4,j}, \quad H_{2,j}=2 J_{3,j}, \\
&E_{1,j}= \frac{1}{\sqrt 2}[J_{1,j}-J_{8,j}+\mathrm{i}(J_{2,j}-J_{7,j})], \quad
E_{2,j} = -(J_{5,j}+\mathrm{i} J_{6,j}), \quad
E_{3,j} = \frac{1}{\sqrt 2}[J_{1,j}+J_{8,j}+\mathrm{i} (J_{2,j}+J_{7,j})], \\ 
&F_{1,j}=\frac{1}{\sqrt 2}[J_{1,j}-J_{8,j}-\mathrm{i} (J_{2,j}-J_{7,j})],\quad
F_{2,j}=-(J_{5,j}-\mathrm{i} J_{6,j}), \quad
F_{3,j}= \frac{1}{\sqrt 2}[J_{1,j}+J_{8,j}-\mathrm{i} (J_{2,j}+J_{7,j})],
\end{eqnarray*}
and for even $j$,
 \begin{eqnarray*}
&H_{1,j}= J_{3,j}+J_{4,j}, \quad H_{2,j}= 2J_{3,j},\\
&E_{1,j}= \frac{1}{\sqrt 2}[J_{1,j}-J_{8,j}+\mathrm{i} (J_{2,j}-J_{7,j})],\quad
E_{2,j}= -(J_{5,j}+\mathrm{i} J_{6,j}),\quad
E_{3,j}= \frac{1}{\sqrt 2}[J_{1,j}+J_{8,j}+\mathrm{i} (J_{2,j}+J_{7,j})],\\
&F_{1,j}= \frac{1}{\sqrt 2}[J_{1,j}-J_{8,j}-\mathrm{i} (J_{2,j}-J_{7,j})], \quad
F_{2,j}=-(J_{5,j}-iJ_{6,j}), \quad
F_{3,j}= \frac{1}{\sqrt 2}[J_{1,j}+J_{8,j}-\mathrm{i} (J_{2,j}+J_{7,j})].
\end{eqnarray*}

The following relations are satisfied
\begin{eqnarray} \label{eqnsA2}
&[H_1,E_1]=2E_1, \quad [H_1,F_1]=-2F_1,\quad [E_1,F_1]=H_1,\quad 
[H_2,E_2]=2E_2,\nonumber\\
&[H_2,F_2]=-2F_2,\quad [E_2,F_2]=H_2,\quad [H_2-H_1,E_3]=2E_3, \quad [H_2-H_1,F_3]=-2F_3,\nonumber\\
&[E_3,F_3]=H_2-H_1, \quad [H_1,E_2]=E_2, \quad [H_1,E_3]=-E_3,\quad
[H_2,E_1]=E_1,\\
&[H_2,E_3]=E_3, \quad [H_1,F_2]=-F_2, \quad 
[H_1,F_3]=F_3, \quad [H_2,F_1]=-F_1, \quad [H_2,F_3]=-F_3,\nonumber\\
&[F_1,F_2]=0, \quad [F_2,F_3]=0, \quad [F_1,F_3]=-F_2, \quad [E_1,E_2]=0, \quad [E_2,E_3]=0, \quad [E_1,E_3]=E_2 \,. \nonumber
\end{eqnarray}

Denoting $|1\rangle$, $|0\rangle$ and $|\rm{-}1\rangle$ as the eigenvectors of the spin operator $S^z_{j}$,
with the respective eigenvalues $1$, $0$ and $-1$, we have
$E_1|1...1\rangle=0$, $E_2|1...1\rangle=0$ and $E_3|1...1\rangle=0$.
Therefore, by definition, $|1...1\rangle$ is the highest weight state.

(ii) The second choice for the eight generators, commuting with the Hamiltonian (\ref{hambq}), are again as given in (\ref{8gens}),
where now for odd $j$,
\begin{eqnarray*}
&H_{1,j} = (S^y_{j})^{2}-(S^z_{j})^{2}, \quad H_{2,j} = (S^y_{j})^{2}-(S^x_{j})^{2}, \\
&E_{1,j} = S^y_{j}\,S^z_{j}, \quad E_{2,j} = S^y_{j}\,S^x_{j}, \quad E_{3,j}\ = S^z_{j}\,S^x_{j}, \\
&F_{1,j} = S^z_{j}\,S^y_{j}, \quad F_{2,j} = S^x_{j}\,S^y_{j}, \quad F_{3,j} = S^x_{j}\,S^z_{j},
\end{eqnarray*}
and for even $j$,
\begin{eqnarray*}
&H_{1,j} = (S^z_{j})^{2}-(S^y_{j})^{2}, \quad H_{2,j} = (S^x_{j})^{2}-(S^y_{j})^{2},\\
&E_{1,j} = S^z_{j} \, S^y_{j}, \quad E_{2,j} = S^x_{j} \, S^y_{j}, \quad E_{3,j} = S^x_{j} \, S^z_{j},\\
&F_{1,j} = S^y_{j} \, S^z_{j}, \quad F_{2,j} = S^y_{j} \, S^x_{j}, \quad F_{3,j} = S^z_{j} \, S^x_{j}.
\end{eqnarray*}
They satisfy the same relations~(\ref{eqnsA2}).

In addition, the local components of the  eight generators for the staggered symmetry group ${\rm SU}(3)$ may be expressed in terms of $J_{\eta,j}$, $\eta = 1,\ldots, 8$.
For odd $j$, we have
\begin{eqnarray*}
&H_{1,j}= -J_{4,j}+J_{5,j}, \quad H_{2,j}=2J_{5,j}, \\
&E_{1,j}= \mathrm{i} J_{1,j}- J_{7,j}, \quad E_{2,j}= -\mathrm{i} J_{3,j}- J_{6,j}, \quad 
E_{3,j}= \mathrm{i} J_{2,j}- J_{8,j}, \\
& F_{1,j}= - \mathrm{i} J_{1,j}- J_{7,j}, \quad F_{2,j}= \mathrm{i} J_{3,j}- J_{6,j}, \quad 
F_{3,j}= -\mathrm{i} J_{2,j}- J_{8,j},
\end{eqnarray*}
and for even $j$,
 \begin{eqnarray*}
& H_{1,j}= -J_{4,j}+ J_{5,j}, \quad H_{2,j}=2J_{5,j}, \\
& E_{1,j}= -\mathrm{i} J_{1,j}+ J_{7,j}, \quad E_{2,j}= \mathrm{i} J_{3,j}+ J_{6,j}, \quad 
E_{3,j}= - \mathrm{i} J_{2,j}+ J_{8,j}, \\
& F_{1,j}=  \mathrm{i} J_{1,j}+ J_{7,j}, \quad F_{2,j}= -\mathrm{i} J_{3,j}+ J_{6,j},\quad 
F_{3,j}= \mathrm{i} J_{2,j}+ J_{8,j}.
\end{eqnarray*}

Denoting $|0_x\rangle$ and $|0_y\rangle$ as the eigenvectors of the spin operators $S^x_{j}$ and $S^y_{j}$, corresponding to the respective eigenvalues $0$, we have $E_1|0_x0_y...0_x0_y\rangle=0$, $E_2|0_x0_y...0_x0_y\rangle=0$ and $E_3|0_x0_y...0_x0_y\rangle=0$.
Therefore, by definition, $|0_x0_y...0_x0_y\rangle$ is the highest weight state.

We remark that the two choices for the highest weight state are unitarily equivalent to each other, under a local unitary transformation $U$, satisfying $K_x=UK_yU^{\dagger}$,  $K_y=UK_zU^{\dagger}$ and  $K_z=UK_xU^{\dagger}$, with $K_x=J_5$, $K_y=J_6$ and $K_z=J_3$ being the three generators of a symmetry subgroup ${\rm SU}(2)$.
Here, $U$ takes a product form $\prod_jU_j$, where
\begin{equation} \label{eqnsA3}
U_j=A+B \, S^z_{j}+C\, (S^z_{j})^2+D[(S^x_{j})^2-(S^y_{j})^2+S^x_{j} \, S^y_{j} + S^y_{j} \, S^x_{j}],
\end{equation}
with $A=\mathrm{i}$, $B=(-1)^j(-1+\mathrm{i})\sqrt{2}/4$, $C=(-1)^j\sqrt{2}/4+[(-1)^j\sqrt{2}-4]\mathrm{i}/4$, and
$D=(-1+\mathrm{i})\sqrt{2}/4$.

\subsection{Fibonacci-Lucas sequences in the ferromagnetic spin-1 biquadratic model}
\label{AppB}
\renewcommand{\theequation}{B\arabic{equation}}
\renewcommand{\thefigure}{B\arabic{figure}}
\setcounter{equation}{0}
\setcounter{figure}{0}
The two well-known Fibonacci-Lucas sequences are characterized by means of the characteristic equation of the form $X^2-PX+Q=0$, 
when $P$ and $Q$ are co-prime integers, i.e., $(P,Q)=1$~\cite{Fibonacci}.
If $u$ and $v$ are the roots of the characteristic equation, then $u+v=P$ and $uv=Q$.
This leads to the  two Fibonacci-Lucas sequences $U_n(P,Q)$ and $V_n(P,Q)$, with
\begin{eqnarray}
	U_n(P,Q)=\frac{u^n-v^n}{u-v},\qquad 
	V_n(P,Q)=u^n+v^n,
\end{eqnarray}
where $n$ is a non-negative integer.
The sequences satisfy the three-term recursive relations $U_{n+2}(P,Q)=PU_{n+1}(P,Q)-QU_n(P,Q)$, with $U_0(P,Q)=0$,  $U_1(P,Q)=1$ and 
$V_{n+2}(P,Q)=PV_{n+1}(P,Q)-QV_n(P,Q)$, with $V_0(P,Q)=2$ and $V_1(P,Q)=P$. 
The key feature is that such sequences exhibit self-similarities, since it is possible to build self-similar rectangles, 
with the ratio $r=\sqrt{u}/\sqrt{v}$ between their length and width, as an extension of the well-known golden rectangles.

\begin{table}[htbp]
	\centering
	\label{tab1}
	\begin{tabular}{|c|c|c|}
		\hline
		$L$& ${\rm dim }(\Omega_L^{\rm OBC})$& ${\rm dim }(\Omega_L^{\rm PBC})$\\
       \hline
        2& 8&8\\
		3 & 21&18\\
		4 & 55&47\\
		5 & 144&123\\
		6 & 377&322\\
		7 & 987&843\\
		8 & 2584&2207\\
		10 & 17711&15127\\
		\hline
	\end{tabular}
    	\caption{Ground state degeneracies ${\rm dim }(\Omega_L^{\rm OBC})$  and  ${\rm dim }(\Omega_L^{\rm PBC})$ under OBCs and PBCs with increasing size $L$ for the ferromagnetic spin-1 biquadratic model (3). For each case the ground state energy per site $E_{gs}^{\rm OBC}(L)=E_{gs}^{\rm PBC}(L)=1$.}
\label{TableC}
\end{table}

It has been argued in Ref.~\cite{shiqianqian} that the highly degenerate ground states arising from SSB with type-B GMs admit an exact Schmidt decomposition, thus exhibiting self-similarities as a whole.
In this way a fractal structure underlies the ground state subspace.
If we choose the ground state degeneracies ${\rm dim }(\Omega_L^{\rm OBC})$ and ${\rm dim }(\Omega_L^{\rm PBC})$, or equivalently, the residual entropy per site, as a physical observable, one may anticipate that they exhibit such self-similarities.
Given these quantities are integers, we have to cope with a situation to express an integer in terms of an irrational number, as encountered in the Binet formula~\cite{binet}.
A tentative speculation is that the ground state degeneracies ${\rm dim }(\Omega_L^{\rm OBC})$ and ${\rm dim }(\Omega_L^{\rm PBC})$ satisfy a three-term recursive
relation, as occurs in the Fibonacci-Lucas sequences.
A simple calculation for small $L$ values, as shown in Table~\ref{TableC}, demonstrates that the ground state degeneracies  ${\rm dim }(\Omega_L^{\rm OBC})$ and ${\rm dim }(\Omega_L^{\rm PBC})$ constitute the two Fibonacci-Lucas sequences, with values $P=3$ and $Q=1$.
In fact, the three-term recursive relations for the Fibonacci-Lucas sequences may be recognized to be identical to those for the ground state degeneracies  ${\rm dim }(\Omega_L^{\rm OBC})$  for $L\geq 2$ and ${\rm dim }(\Omega_L^{\rm PBC})$ for $L\geq 3$, if $n=L+1$ for OBCs and $n=L$ for PBCs.
As such, we reproduce the results obtained by Aufgebauer and Kl\"{u}mper from consideration of the underlying Temperley-Lieb algebra~\cite{spins}.

The results can thus be summarised as
\begin{align}
	{\rm dim }(\Omega_L^{\rm OBC}) &= \frac{(3+\sqrt 5)^{L+1} - (3-\sqrt 5)^{L+1}}{ 2^{L+1} \sqrt{5}  },\\
	{\rm dim }(\Omega_L^{\rm PBC}) &= \frac{(3+\sqrt 5)^L + (3-\sqrt 5)^L}{ 2^L },
\end{align}
for $L \ge 2$ and for $L \ge 3$, respectively.
The ground state degeneracies have also been recognized as the Fibonacci numbers in Ref.~\cite{saleur} (see also Ref.~\cite{moudgalya})
and the bisection of the Lucas numbers~\cite{katsura}.
\footnote{According to Ref.~\cite{saleur} this observation goes back to F.D.M Haldane and D.J. Arovas in private communication.}
Here we have established the connection to self-similarities as a reflection of an abstract fractal underlying the ground state subspace.

\subsection{The set for all the fillings of atypical generalized highest weight states is dense}\label{AppC}
 \renewcommand{\theequation}{C\arabic{equation}}
 \renewcommand{\thefigure}{C\arabic{figure}}
 \setcounter{equation}{0}
 \setcounter{figure}{0}
As mentioned in the main text, a hierarchical structure emerges behind all exponentially many degenerate ground states, with atypical generalized highest weight states being the most fundamental. As defined,  for the model under investigation, an atypical generalized highest weight state may be constructed in terms of local states $|0\rangle$ and $|1\rangle$ that are eigenvectors of $S_j^z$ with eigenvalues 0 and 1, respectively. 
Here we define the filling $f_0=N_0/L$ for a generalized highest weight state, where $N_0$ counts the number of local states $|0\rangle$ in a specific generalized highest weight state. Given the periodicity of an atypical generalized highest weight  state, its filling becomes $f_0=N_{p0}/p$, with $N_{p0}$ being the number of local states $|0\rangle$ in one {\it emergent} unit cell, consisting of $p$ adjacent lattice sites.
Note that the filling  defined in the main text is relative to the highest weight state or a chosen generalized highest weight state.

As it turns out, the set of all the fillings of atypical generalized highest weight states forms a dense subset with respect to all possible fillings, if we are only interested in this set in the thermodynamic limit.  Indeed, $p$ take any integer, as already mentioned in the main text.

Consider two atypical generalized highest weight states with the periods $p_1$ and $p_2$, with their fillings being $f_{0,1}=N_{p0,1}/p_1$ and $f_{0,2}=N_{p0,2}/p_2$, respectively. Without loss of generality, we assume that $f_{0,1} < f_{0,2}$. It is readily seen that the filling $f_0=N_{p0}/p$, with $p=2p_1 p_2$ and $N_{p0}=N_{p0,1} p_2 +N_{p0,2} p_1$, satisfies the condition 
$f_{0,1} < f_0 < f_{0,2}$, given $f_0 = 1/2(f_{0,1}+f_{0,2})$.
In other words, for any two given fillings  $f_{0,1}$ and  $f_{0,2}$ for two atypical generalized highest weight states, no matter how close they are, there is always one filling for another atypical generalized highest weight state that lies between them. Hence the set of all the fillings of atypical generalized highest weight states is dense. 

Given that all atypical degenerate ground states are generated from the action of the lowering operators on all atypical generalized highest weight states, we are led to conclude that the set for all the fillings of atypical degenerate ground states is dense with respect to all possible fillings in the thermodynamic limit.

\subsection{Scaling of the entanglement entropy in the thermodynamic limit from a finite-size approach}
\label{AppD}
\renewcommand{\theequation}{D\arabic{equation}}
\renewcommand{\thefigure}{D\arabic{figure}}
\setcounter{equation}{0}
\setcounter{figure}{0}

\begin{figure}[h!]
	\centering
	\includegraphics[width=0.85\textwidth]{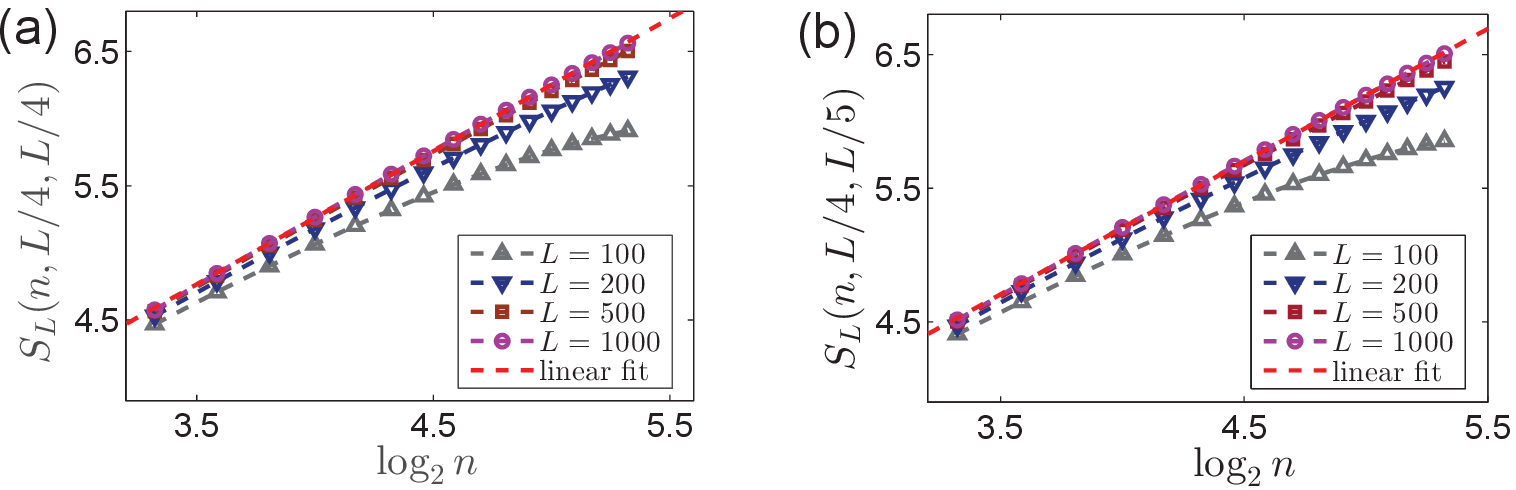}
	\caption{ (a) The entanglement entropy  $S_{\!L}(n,M_1,M_2)$  vs $\log_2 n$ for the ferromagnetic spin-$1$ biquadratic model. Here $M_1=M_2=L/4$ for the indicated $L$ values.
		(b) The entanglement entropy $S_{\!L}(n,M_1,M_3)$ vs $\log_2 n$.
		Here $M_1=L/4$ and $M_3=L/5$ for the indicated $L$ values.
		The block size $n$  is a multiple of two, and ranges from $10$ to $40$.	}
	\label{Su3finitehws}
\end{figure}

\begin{figure}[h!]
	\centering
	\includegraphics[width=0.85\textwidth]{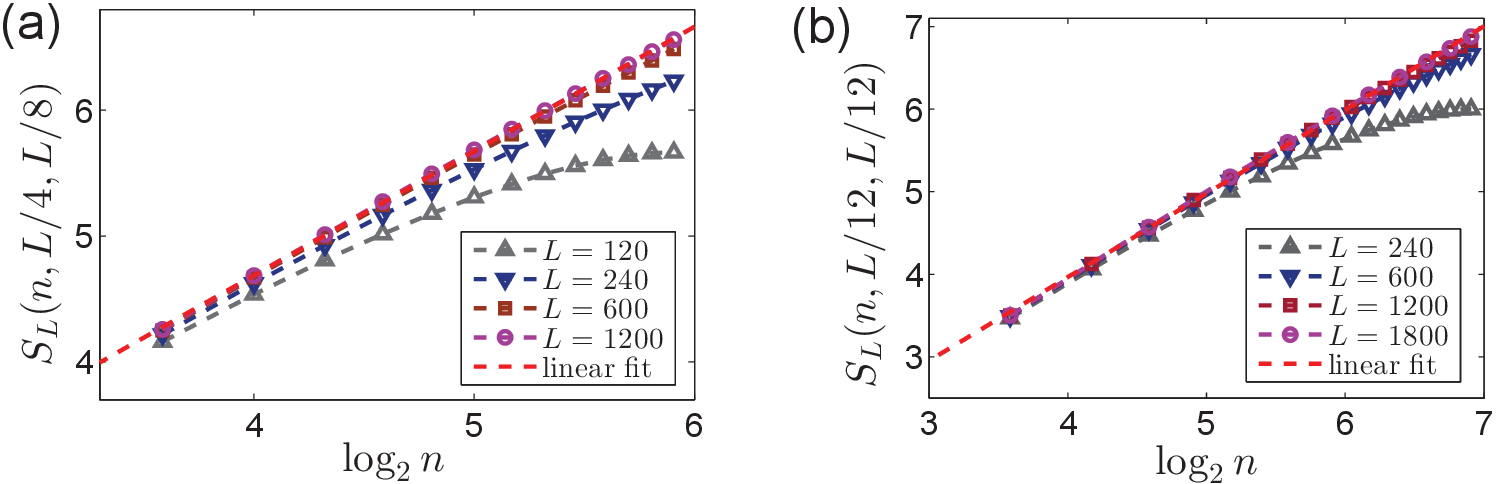}
	\caption{ (a) The entanglement entropy $S_{\!L}(n,M_2,M_3)$ vs $\log_2 n$.
		Here $M_2=L/4$ and $M_3=L/8$  for the indicated $L$ values.
		A generalized highest weight state is chosen to be $|1110...1110\rangle$.
		The block size $n$ is a multiple of four, and ranges from $12$ to $60$.
		(b) The entanglement entropy $S_{\!L}(n,M_1,M_2)$ vs $\log_2 n$.
		Here $M_1=L/12$ and $M_2=L/12$ for the indicated $L$ values.
		A generalized highest weight state is chosen to be $|10\rm{-}1\rm{-}101...10\rm{-}1\rm{-}101\rangle$.
		The block size $n$ is a multiple of six, and ranges from $12$ to $120$.
	}
	\label{Su3finitege}
\end{figure}

To confirm the emergence of the logarithmic scaling behavior of the entanglement entropy in the thermodynamic limit, we show a number of plots in Fig.~\ref{Su3finitehws} and in Fig.~\ref{Su3finitege} for degenerate ground states generated from the highest weight state and generalized highest weight states, respectively.

To begin, in Fig.~\ref{Su3finitehws}(a) we plot the entanglement entropy $S_{\!L}(n,M_1,M_2)$ vs $\log_2 n$ for $|L,M_1,M_2\rangle_2$ given in Eq.~(\ref{lm1m2}).
Here $M_1=M_2=L/4$ for the indicated $L$ values.
A significant deviation from the logarithmic scaling behavior is observed when $L$ is relatively small, but tends to vanish as $L$ increases.
Indeed, the prefactor is close to the exact value $1$, with an error less than 1$\%$ when $L=1000$.
Here we have $S_{\!L}(n,M_1,M_2)=0.991\log_2n+1.298$.

Fig.~\ref{Su3finitehws}(b) shows a plot of the entanglement entropy $S_{\!L}(n,M_1,M_3)$ vs $\log_2 n$ for $|L,M_1,M_3\rangle_2$ given in Eq.~(\ref{lm1m3}).
Here $M_1=L/12$ and $M_2=L/12$ for the indicated $L$ values.
A significant deviation from the logarithmic scaling behavior is again observed when $L$ is relatively small, but tends to vanish as $L$ increases.
The prefactor is again close to the exact value $1$, with an error being less than 1$\%$ when $L=1000$. Here $S_{\!L}(n,M_1,M_3)=0.993\log_2n+1.203$.

The entanglement entropy $S_{\!L}(n,M_2,M_3)$ vs $\log_2 n$ for $|L,M_2,M_3\rangle_4$, as given in Eq.~(\ref{lm2m3p4}), is shown in Fig.~\ref{Su3finitege}(a).
A significant deviation from the logarithmic scaling behavior is again observed when $L$ is relatively small, but tends to vanish as $L$ increases.
Again the prefactor is close to the exact value $1$, with an error being less than 1.5$\%$ when $L=1200$.
Here  $S_{\!L}(n,M_2,M_3)=0.988\log_2n+0.735$.
Fig.~\ref{Su3finitege}(b) shows the plot of the entanglement entropy $S_{\!L}(n,M_1,M_2)$ vs $\log_2 n$ for $|L,M_1,M_2\rangle_6$, as presented in Eq.~(\ref{lm1m2p6}).
A significant deviation from the logarithmic scaling behavior is again observed when $L$ is relatively small, but tends to vanish as $L$ increases.
Here also the prefactor is close to the exact value $1$, with an error being less than 1.5$\%$, when $L=1800$. Here $S_{\!L}(n,M_1,M_2)=1.013\log_2n-0.085$.

\subsection{An over-complete set of spin or generalized coherent states as  (factorized) degenerate ground states}\label{AppE}
\renewcommand{\theequation}{E\arabic{equation}}
\renewcommand{\thefigure}{E\arabic{figure}}
\setcounter{equation}{0}
\setcounter{figure}{0}

To identify the fractal dimension with the number of GMs for the orthonormal basis states in an irreducible representation space of the symmetry group, we resort to an alternative approach developed in our previous work~\cite{cantorset}. For this purpose, we need to introduce spin or generalized coherent states that have been adapted from the conventional definition~\cite{gilmore} to be suitable for our discussion about SSB in the spin-1 biquadratic model, since they may be expanded in terms of generalized highest weight states. As an over-complete set, one may form a linear combination of these spin or generalized coherent states on any fractal $C$ in the coset space, with a Cantor set as a typical example. 
In Ref.~\cite{cantorset}, we have shown that the entanglement entropy $S(L,n)$ for a linear combination of  generalized coherent states on a fractal $C$ scales as 
\begin{equation}
S(L,n)=\frac{d_f}{2}\log_2{\frac{n(L-n)}{n}}+S_0.\label{dfsln} 
\end{equation}
Here $S_0$ is an additive model-dependent constant and $d_f$ is the fractal dimension of a support $C$,  on which a linear combination is formed. Indeed, any support may be approximated in terms of a fractal decomposable into a set of the Cantor sets~\cite{cantorset}.  Note that  
$S(L,n)$ satisfies the inversion symmetry relation $S(L,n)=S(L,L-n)$, as follows from the site-centered inversion symmetry for odd $L$'s and the bond-centered inversion symmetry for even $L$'s.

Here we recall that a Cantor set is created by dividing the interval $[0,\;  1]$ into $1/r$ parts and removing  $1/r -N$ subintervals such that we have $C[N,r;1]$ at step 1. Here, we have assumed that $1/r$ is always a positive integer and $1/r > N$. Repeating the same  procedure for the $N$ remaining subintervals and removing  the subsubintervals in the same way as done for $C[2,1/3;\{k\}]$, we have $C[N,r;2]$ at step 2. Keep repeating the same procedure $k$ times, we have $C[N,r;k]$ at step $k$, thus yielding a self-similar Cantor set, denoted as $C[N,r;\{k\}]$.  Indeed, the number of subintervals in a Cantor set $C[N,r;\{k\}]$ is $N^k$ at the step $k$, so the fractal dimension $d_f$ for the Cantor set $C[N,r;\{k\}]$ is $d_f= -\ln N/\ln r$.

To proceed, it is necessary to distinguish between odd and even $L$'s under PBCs, due to the presence of the odd-even parity effect. We remark that, although we only focus on PBCs,  the discussion below for even $L$'s is  valid  under OBCs, since the symmetry group is the same staggered $\rm{SU(3)}$ group.

\subsubsection{Scaling of the entanglement entropy under PBCs: odd $L$'s}

As already pointed out in Section~\ref{Sec3},  if $L$ is odd, then the model possesses the symmetry group $\rm{SU(2)}$ generated by $S^x$, $S^y$, and $S^z$, where 
$S^x=\sum_jS^x_j$, $S^y=\sum_jS^y_j$ and $S^z=\sum_j S^z_j$. 
The coset space is $CP^{1} \sim \rm{SU(2)}/ \rm{U(1)}$, diffeomorphic to the sphere $S^2$. It accommodates spin coherent states $|\psi(\theta,\phi)\rangle$ that  are  expressed in terms of the two spherical coordinates $\theta \in [0,\pi]$ and $\phi \in [0,2\pi]$ on the sphere $S^2$,
\begin{equation*}
|\psi(\theta,\phi)\rangle=|v(\theta,\phi)\rangle_1 \cdots |v(\theta,\phi)\rangle_j \cdots |v(\theta,\phi)\rangle_L,
\end{equation*}
with
\begin{equation*}
|v(\theta,\phi)\rangle_j=\exp(\mathrm{i}\,\phi S^z_{j})\exp(\mathrm{i}\,\theta S^y_{j})\;|1\rangle_j,
\end{equation*}
where $|1\rangle_j$ represents the eigenvector of $S_j^z$ with eigenvalue $1$ at lattice site $j$. Indeed, one may form a linear combination of a set of the overcomplete generalized coherent states $|\psi(\theta,\phi)\rangle$ on a fractal $C$, with  the Cantor set $C[N,r;\{k\}]$ being a typical example.
Here we adopt a convention that  $C[N,r;\{k\}]$ should be understood as the image on the sphere $S^2$ under the mapping $\phi: [0,1] \rightarrow S^1$, defined as $\phi (\xi) = 2\pi \xi$, or under the mapping $\theta: [0,1] \rightarrow S^1$, defined as $\theta (\xi) = \pi/2 \xi$, where $\xi \in [0,1]$. Owing to a symmetric consideration, the two mappings defined above are sufficient to meet our needs. From now on, for berivity we do not make any distinction between the image of a fractal under the mapping $\phi$ or $\theta$ and  a fractal itself.

As a result of the SSB pattern from the  $\rm{SU(2)}$ symmetry group to the residual symmetry group $\rm{U(1)}$, highly degenerate ground states $|L,M\rangle$, constructed from the repeated action of the lowering operator $S_-$ on the highest weight state $|1\ldots 1\rangle$:
$|L,M\rangle=1/Z(L,M) S_-^M |1\ldots 1\rangle$ ($M=0$, $1$, $2$, \ldots, $2L$),  constitute the orthonormal basis states in the irreducible representation space within the ground state subspace. Here we point out that this irreducible representation space is formally identical to the spin-1 ferromagnetic Heisenberg model~\cite{cantorset}. However, we still reproduce them below for completeness.

The orthonormal basis states $|L,M\rangle$ admit an exact Schmidt decomposition~\cite{shiqianqian}
\begin{equation}
|L,M\rangle= \sum\limits_{\kappa}\lambda(L,n,\kappa,M)
|n,\kappa\rangle|L-n,M-\kappa\rangle. \label{schmidt}
\end{equation}
Here, $\lambda(L,n,\kappa,M)$ denote the Schmidt coefficients, which take the form:
\begin{equation*}
\lambda(L,n,\kappa,M)=\frac{\mu(L,n,\kappa,M)}{\nu(L,n,\kappa,M)},
\end{equation*}
with
\begin{equation*}
\mu(L,n,\kappa,M)\!=\!\sqrt{\!{\sum}'_{n_{-1},\;n_{0},\; n_{1},\atop l_{-1},\;l_{0},\;l_{1}}\!\prod_{u,t=-1}^{0}\!\varepsilon(u)^{n_{u}}
	{C_{n\!-\!\sum_{m=-1}^{u-1}\!n_m}^{n_u}}\!\varepsilon(t)^{l_{t}}{C_{L\!-\!n-\!\sum_{m=-1}^{t-1}l_m}^{l_t}}},
\end{equation*}
and
\begin{equation*}
\nu(L,n,\kappa,M)=\sqrt{{\sum}'_{N_{-1},N_{0},N_{1}}\prod_{u=-1}^{0}\varepsilon(u)^{N_{u}}
	{C_{L-\sum_{m=-1}^{u-1}N_m}^{N_u}}}.
\end{equation*}
The sum $\sum'_{n_{-1},\;n_{0},\;n_{1}}$ is taken over all the possible values of $n_{-1}$, $n_{0}$, $n_1$, subject to the constraints $\sum_{m=-1}^1 n_m=n$ and $\sum_{m=-1}^{1}(1-m)n_m=\kappa$, and the sum $\sum'_{l_{-1},\;l_{0},\;l_{1}}$ is taken over all the possible values of $l_{-1}$, $l_{0}$, $l_1$, subject to the constraints $\sum_{m=-1}^1 l_m=L-n$ and $\sum_{m=-1}^{1}(1-m)l_m=M-\kappa$.
Here $C_{M}^{\kappa}$ is the binomial coefficient: $C_{M}^{\kappa}= M !/ (\kappa ! (M - \kappa)!)$ and $\varepsilon(u)$ is given by
\begin{equation*}
\varepsilon(u)=\frac{\prod_{m=u+1}^{1}{(1+m)(1-m+1)}}{\prod_{m=u}^{0}(1-m)^2}. \label{epsilon}
\end{equation*}

\begin{figure}[h!]
	\centering
	\includegraphics[width=0.6\textwidth]{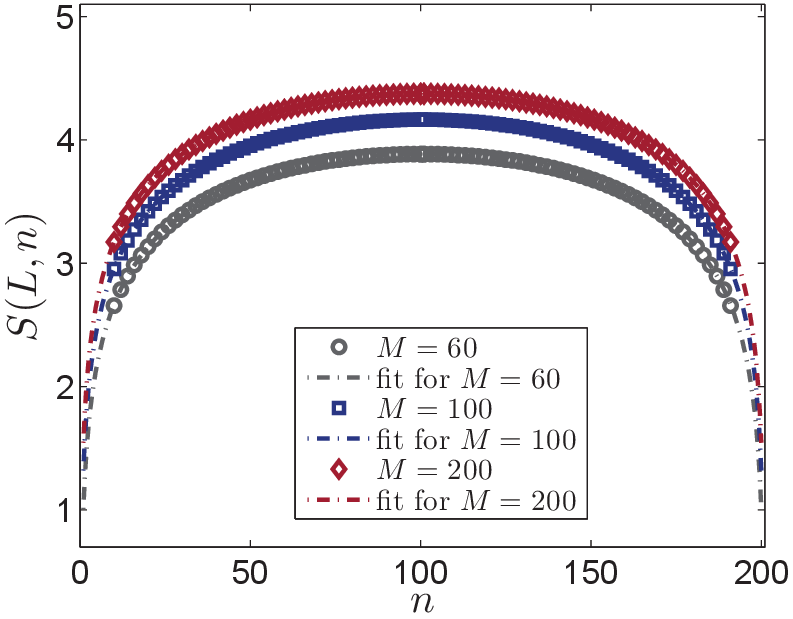}
	\caption{The entanglement entropy $S(L,n)$ versus $n$ for  the orthonormal basis state $|L,M\rangle$ in the irreducible representation space of the symmetry group $\rm{SU(2)}$ for odd $L$'s, where $L=201$, with $M=50$, $M=100$ and $M=200$, respectively. The prefactor is half the number of GMs $N_B=1$, with a relative error being less than 2$\%$.}
	\label{A1}
\end{figure}

We plot the entanglement entropy $S(L,n)$ versus $n$ for different value of $M$ with $L=201$ in Fig.~\ref{A1}. 
The prefactor is half the type-B Goldstone mode $N_B=1$, with a relative error being less than 2$\%$.

\begin{figure}[h!]
	\centering
	\includegraphics[width=0.6\textwidth]{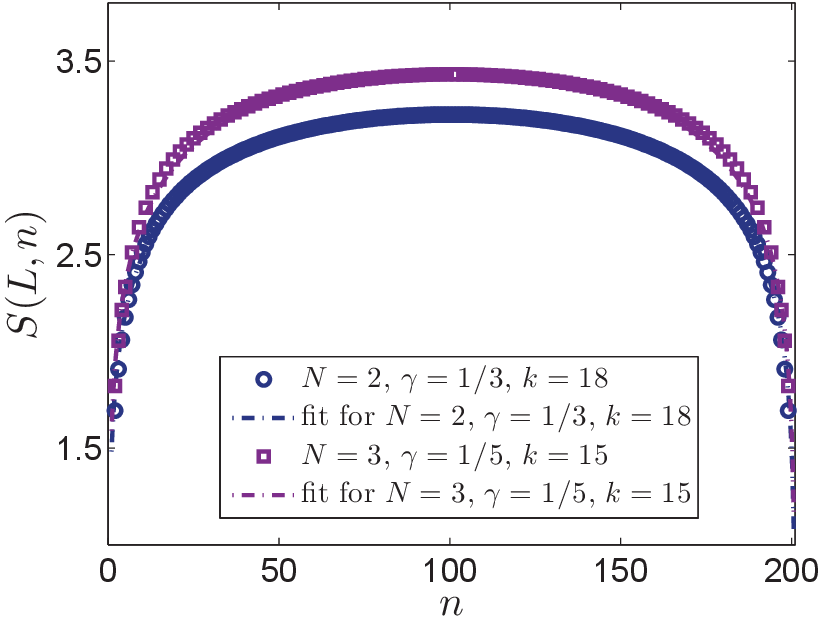}
	\caption{The entanglement entropy $S(L,n)$ versus $n$ for a linear combination of spin coherent states as a degenerate ground state $|\Phi_C(\theta)\rangle$ on the Cantor set $C[N,r;\{k\}]$ with $\theta=\pi/2$ in the spin-$1$ biquadratic model, with equal coefficients for $N=2$ and $r=1/3$ and  $N=3$ and $r=1/5$. The prefactor is half the fractal dimension $d_f$ of the Cantor set $C[N,r;\{k\}]$: $d_f=-\ln N/\ln r$, with a relative error being less than 3.5$\%$.}
	\label{A2}
\end{figure}

Note that $|\psi(\theta,\phi)\rangle$ may be expanded into a linear combination
in terms of $|L,M\rangle$: $|\psi(\theta,\phi)\rangle=\sum_{M=0}^{2L}a_{LM}(\theta,\phi)|L,M\rangle$, where $a_{LM}$ are complex numbers, which are formally equal to $a_{LM}(\theta,\phi)= \langle L,M |\psi(\theta,\phi)\rangle$. As it turns out, $a_{LM}(\theta,\phi)=b_{LM}(\theta)\exp(i\phi(L-M))$. Here we present the explicit expression for $b_{LM}(\theta)$
\begin{equation*}
b_{LM}(\theta)=\frac{(-1)^M(\cos \theta+1)^{L-M}\sin \theta^M}{2^{L-\frac{M}{2}}Z(L,M)}
\sum_{j=0}^{M/2}2^{-2j}C_L^jC_{L-j,M-2j}.
\end{equation*}
Taking advantage of $1/(2\pi)\int_{0}^{2\pi} d\phi  \exp(i\phi(L-M)) = \delta_{L\;M} $, we are able to express $|L,M\rangle$  ($M=0$, \ldots, $2L$) in terms of $|\psi(\theta,\phi)\rangle$ on a circle $S^1$ with fixed $\theta$ ($0<\theta<\pi$),
\begin{equation}
|L,M\rangle=\frac {1} {b_{LM}(\theta)}\int_{0}^{2\pi} d\phi \; \exp \left(-\mathrm{i}\phi(sL-M)\right)|\psi(\theta,\phi)\rangle.
\label{lmexp}
\end{equation}
Obviously, this representation may be regarded as a linear combination in the overcomplete basis states $|\psi(\theta,\phi)\rangle$, with the coefficients only involving a phase factor.  
Hence  the fractal dimension $d_f$ is equal to one for the orthonormal basis states $|L,M\rangle$, identical to the number of type-B GMs for odd $L$'s.

We turn to a  linear combination of spin coherent states $|\Phi_C(\theta)\rangle$, which takes the form
\begin{equation}
|\Phi_C(\theta)\rangle=\frac{1}{Z_C}\sum_{\phi_\gamma \in C} c(\phi_\gamma) |\psi(\theta,\phi_\gamma)\rangle, \label{lcfractal}
\end{equation}
where $c(\phi_\gamma)$ ($\gamma =1,2,\ldots,|C|$) are complex numbers, with $|C|=N^k$ being the number of the subintervals in  the Cantor set $C[N,r;\{k\}]$, and ${Z_C}$ is a normalization factor to ensure that $|\Phi_C(\theta)\rangle$ has been normalized, and the sum over $\phi_\gamma$ is carried out for all the subintervals at the step $k$. As for restrictions  imposed on the coefficients $c(\phi_\gamma)$ in the linear combinations, we refer to Ref.~\cite{cantorset}. Here we only consider a linear combination with equal coefficients for brevity.
Shown in Fig.~\ref{A2} is the entanglement entropy $S(L,n)$  versus $n$  for such a linear combination  $|\Phi_C(\theta)\rangle$ on the Cantor set $C[N,r;\{k\}]$ with $\theta=\pi/2$ for odd $L$'s, when $L=201$, with  $N=2$, $r=1/3$ at the step $k=18$, and  $N=3$, $r=1/5$ at the step $k=15$. 
As it turns out, the prefactor in Eq.~(\ref{dfsln}) is half the fractal dimension $d_f$ of the Cantor set $C[N,r;\{k\}]$: $d_f=-\ln N/\ln r$, with a relative error being less than 2$\%$.

\subsubsection{Scaling of the entanglement entropy under PBCs:  even $L$'s}

If $L$ is even, then the symmetry group becomes  the staggered ${\rm SU}(3)$ group. 
One may introduce two ${\rm SU}(2)$ subgroups, each of which is associated with one of the two type-B GMs as a result of SSB.
We denote the three generators for each of the two ${\rm SU}(2)$  subgroups as $\Sigma_{\alpha}^x$, $\Sigma_{\alpha}^y$ and $\Sigma_{\alpha}^z$ ($\alpha=1,2$). They are defined as $\Sigma_{\alpha,j}^x=(E_{\alpha,j}+F_{\alpha,j})/2$, $\Sigma_{\alpha,j}^y=-i(E_{\alpha,j}-F_{\alpha,j})/2$ and  $\Sigma_{\alpha,j}^z=H_{\alpha,j}/2$. The generators $\Sigma_{\alpha}^x$, $\Sigma_{\alpha}^y$ and $\Sigma_{\alpha}^z$ satisfy the commutation relations: $[\Sigma_{\alpha}^{x},\Sigma_{\alpha}^{y}]=i\Sigma_{\alpha}^{z}$, $[\Sigma_{\alpha}^{y},\Sigma_{\alpha}^{z}]=i\Sigma_{\alpha}^{x}$ and $[\Sigma_{\alpha}^{z},\Sigma_{\alpha}^{x}]=i\Sigma_{\alpha}^{y}$ ($\alpha=1,2$).
An overcomplete set of generalized coherent states $|\psi(\theta_1,\phi_1;\theta_{2},\phi_{2})\rangle$ may be parameterized in terms of  $\theta_\alpha \in [0,\pi]$ and $\phi_\alpha \in [0,2\pi]$ ($\alpha=1 $ and $2$) as follows
	\begin{equation*}
	|\psi(\theta_1,\phi_1;\theta_{2},\phi_{2})\rangle=|v(\theta_1,\phi_1;\theta_{2},\phi_{2})\rangle_1 \cdots |v(\theta_1,\phi_1;\theta_{2},\phi_{2})\rangle_{2l-1} |v(\theta_1,\phi_1;\theta_{2},\phi_{2})\rangle_{2l}\cdots |v(\theta_1,\phi_1;\theta_{2},\phi_{2})\rangle_L,
	\end{equation*}
with
	\begin{equation*}
	|v(\theta_1,\phi_1;\theta_{2},\phi_{2})\rangle_{2l-1/2l}=\exp(\mathrm{i}\phi_{2} \Sigma_{2,2l-1/2l}^{z})\exp(\mathrm{i}\theta_{2} \Sigma_{2,2l-1/2l}^{y})\exp(\mathrm{i}\phi_1 \Sigma_{1,2l-1/2l}^{z})\exp(\mathrm{i}\theta_1 \Sigma_{1,2l-1/2l}^{y})\;|1\rangle_{2l-1/2l}.
	\end{equation*}
Here $|1\rangle_{2l-1/2l}$  represents the eigenstate of $S_{2l-1/2l}^z$ with the eigenvalue being $1$ at the lattice site $2l-1/2l$, where $l=1,2,\ldots,L/2$. Note that the coset space, denoted as $CP^{2}_{\pm}$, contains this two-dimensional (complex) manifold. Indeed, $|\psi(\theta_1,\phi_1;\theta_{2},\phi_{2})\rangle$ may be regarded as a variant of the generalized coherent states for the uniform ${\rm SU}(3)$ symmetry group, which has been exploited in Ref.~\cite{cantorset} to investigate the entanglement entropy for the ${\rm SU}(3)$ spin-1 ferromagnetic model, with the coset space being $CP^2$.  
Specifically, at the  $(2l-1)$-th lattice site, we have
\begin{equation*}
		|v(\theta_1,\phi_1;\theta_2,\phi_2)\rangle_{2l-1}=\left (\cos\frac{\theta_1}{2}\cos\frac{\theta_2}{2}\exp(\mathrm{i}\frac{\phi_1 + \phi_2}{2});
		-\sin\frac{\theta_1}{2}\exp(-\mathrm{i}\frac{\phi_1}{2});
		-\cos\frac{\theta_1}{2}\sin\frac{\theta_2}{2}\exp(\mathrm{i}\frac{\phi_1 - \phi_2}{2}) \right )^T,
\end{equation*}
and at the $2l$-th lattice site, we have
\begin{equation*}
		|v(\theta_1,\phi_1;\theta_2,\phi_2)\rangle_{2l}=\left (\cos\frac{\theta_2}{2}\exp(\mathrm{i}\frac{\phi_2}{2});0;\sin\frac{\theta_2}{2}\exp(-\mathrm{i}\frac{\phi_2}{2}) \right )^T.
\end{equation*}
Here $T$ denotes the transpose of a vector.

The orthonormal basis states $|L,M_1,M_2\rangle_2$ in Eq.~(7)  span the irreducible representation space of the staggered ${\rm SU}(3)$ symmetry group within
the ground state subspace.  Hence, $|\psi(\theta_1,\phi_1;\theta_2,\phi_2)\rangle$ may be expanded into a linear combination in terms of $|L,M_1,M_2\rangle_2$:  $|\psi(\theta_1,\phi_1;\theta_2,\phi_2)\rangle=\sum_{M_1=0}^{L/2}\sum_{M_2=0}^{L}a_{LM_1M_2}(\theta_1,\phi_1;\theta_2,\phi_2) |L,M_1,M_2\rangle_2$, where $a_{LM_1M_2}$ are complex numbers, which are formally equal to $a_{LM_1M_2}(\theta_1,\phi_1;\theta_2,\phi_2)= \langle L,M_1,M_2 |\psi(\theta_1,\phi_1;\theta_2,\phi_2)\rangle$. As it turns out, $a_{LM_1M_2}(\theta_1,\phi_1;\theta_2,\phi_2)=b_{LM_1M_2}(\theta_1,\theta_2)\exp(\mathrm{i}\phi_1/2(L/2-2M_1))\exp(\mathrm{i}\phi_2/2(L-M_1-2M_2))$. Here, we present the explicit expression for $b_{LM_1M_2}(\theta_1,\theta_2)$ 
\begin{eqnarray*}
b_{LM_1M_2}(\theta_1,\theta_2)=
(-1)^{M_1+M_2}(\sin\frac{\theta_1}{2})^{M_1}(\cos\frac{\theta_1}{2})^{L/2-M_1}
(\sin\frac{\theta_2}{2})^{M_2}
(\cos\frac{\theta_2}{2})^{L-M_1-M_2}\sqrt{C_{L/2}^{M_1}C_{L-M_1}^{M_2}}.
\end{eqnarray*}

Taking advantage of $1/(2\pi)\int_{0}^{2\pi} d\phi_1 \exp(\mathrm{i}\phi_1(L/4-M_1)) = \delta_{L/4\;M_1} $ and $1/(2\pi)\int_{0}^{2\pi} d\phi_2 \exp(\mathrm{i}\phi_2(L/2-M_1/2-M_2)) = \delta_{L\;M_1+2M_2}$, we are able to express $|L,M_1,M_2\rangle_2$  in terms of $|\psi(\theta_1,\phi_1;\theta_2,\phi_2)\rangle$ on the coset space  $CP^{2}_{\pm}$, with fixed $\theta_1$ and $\theta_2$ ($0<\theta_1<\pi$ and $0<\theta_2<\pi$),
\begin{equation}
|L,M_1,M_2\rangle_2= \frac {1} {b_{LM_1M_2}(\theta_1,\theta_2)}\int_{0}^{2\pi} d\phi_1 \exp(-\mathrm{i}\phi_1(L/4-M_1)) \int_{0}^{2\pi} d\phi_2 \exp(-\mathrm{i}\phi_2(L/2-M_1/2-M_2))|\psi(\theta_1,\phi_1;\theta_2,\phi_2)\rangle.
\label{lmexp}
\end{equation}
Obviously, this representation may be regarded as a linear combination in terms of the overcomplete set $|\psi(\theta_1,\phi_1;\theta_2,\phi_2)\rangle$, with the coefficients only involving a phase factor.  
Hence the fact that the fractal dimension $d_f$ is equal to two for the orthonormal basis states $|L,M_1,M_2\rangle_2$, identical to the number of type-B GMs.

For even $L$, consider a linear combination on a fractal decomposable into a set of the Cantor sets on the circles by setting $\theta_1$ and $\theta_2$ to be certain values in the interval $(0,\pi)$, whereas $\phi_1$ and $\phi_2$ vary from 0 to $2\pi$.
Specifically,  we may consider a linear combination on a fractal that is decomposable into two Cantor sets $C[N_1;r_1;k_1]$ and  $C[N_2;r_2;k_2]$, located on two circles $S^1$'s with fixed  $\theta_1$ and $\theta_2$,
\begin{align*}
|\Phi_C(\theta_1,\theta_{2})\rangle=&\frac{1}{Z_{C}} \sum_{\substack{\phi_{1,\beta} \in C[N_1;r_1;k_1],\\\;\phi_{2,\gamma} \in C[N_2;r_2;k_2]}} c(\phi_{1,\beta},\;\phi_{2,\gamma}) |\psi(\theta_1,\phi_{1,\beta};\theta_{2},\phi_{2,\gamma})\rangle, \label{lcsu3}
\end{align*}
where $c(\phi_{1,\beta},\;\phi_{2,\gamma})$ ($\beta =1,2,\ldots,N_1^{k_1}$, $\gamma =1,2,\ldots,N_2^{k_2}$) are complex numbers, and ${Z_C}$ is a normalization factor to ensure that $|\Phi_C(\theta_1,\theta_{2})\rangle$ has been normalized. For brevity, we  only consider the case with equal coefficients, namely, $c(\phi_{1,\beta},\;\phi_{2,\gamma})=1$.

\begin{figure}[ht!]
	\centering
	\includegraphics[width=0.6\textwidth]{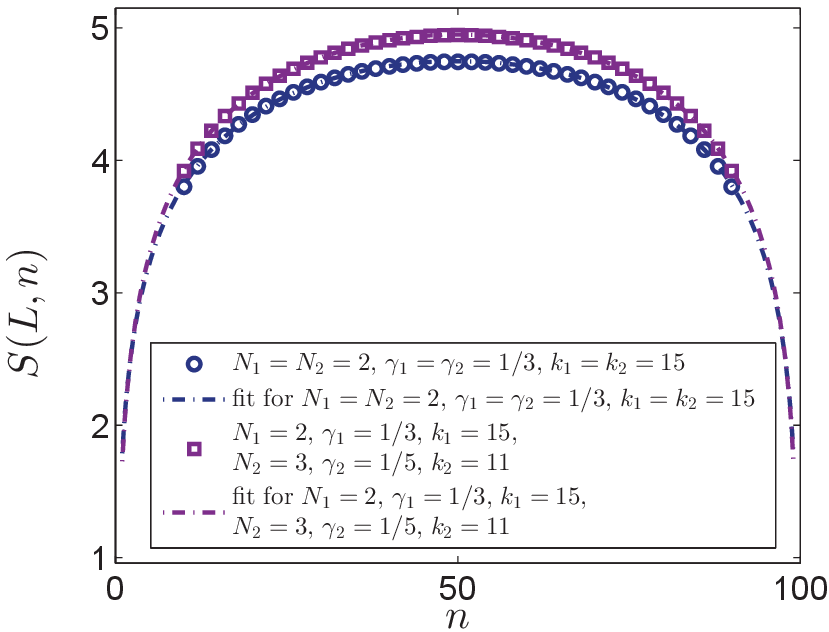}
	\caption{The entanglement entropy $S(L,n)$ versus $n$ for a linear combination of generalized coherent states $|\Phi_C(\theta_1,\theta_{2})\rangle$  on a fractal decomposable into two Cantor sets $C[N_1,r_1;\{k_1\}]$ and $C[N_2,r_2;\{k_2\}]$ with $\theta_1=\theta_2=\pi/2$, for even $L$'s,  when $L=100$, with $N_1=N_2=2$, $r_1=r_2=1/3$ and $k_1=k_2=15$, and $N_1=2$,  $r_1=1/3$, $k_1=15$, $N_2=3$, $r_2=1/5$ and $k_2=11$, respectively.
	The prefactor is half the sum of the fractal dimensions $d_{f1}$ and $d_{f2}$: $d_{f1} = -\ln N_1/\ln r_1$ and $d_{f2} = -\ln N_2/\ln r_2$, with a relative error being less than 1.5$\%$.}
	\label{spinsu3g3g5}
\end{figure}

 In Fig.~\ref{spinsu3g3g5}, we plot the entanglement entropy $S(L,n)$  versus $n$  for such a linear combination $|\Phi_C(\theta_1,\theta_{2})\rangle$, with equal coefficients, on a fractal decomposable into two Cantor sets, with $\theta_1=\theta_2=\pi/2$,  when $L=100$. The Cantor sets are taken to be $C[N_1,r_1;\{k_1\}]$ and $C[N_2,r_2;\{k_2\}]$, with  $N_1=N_2=2$, $r_1=r_2=1/3$ at the step $k_1=k_2=15$, and $N_1=2$,  $r_1=1/3$ at the step $k_1=15$,  $N_2=3$, $r_2=1/5$ at the step $k_2=11$.

As demonstrated in Ref.~\cite{cantorset}, the set of the fractal dimensions for the Cantor sets forms an uncountably infinite subset that is dense for all possible values of the fractal dimension $d_f$. We are thus led to the identification of the fractal dimension with the number of type-B GMs for the orthonormal basis in the irreducible representation space of  the staggered ${\rm SU}(3)$ group for even $L$ under PBCs.

\end{document}


\title{Supplementary Material\\``Goldstone modes and the golden spiral in the ferromagnetic spin-1 biquadratic model"}
	
\author{Huan-Qiang Zhou}
\affiliation{Centre for Modern Physics,
	Chongqing University, Chongqing 400044, The People's Republic of China}

\author{Qian-Qian Shi}
\affiliation{Centre for Modern Physics,
	Chongqing University, Chongqing 400044, The People's Republic of China}

\author{Ian P. McCulloch}
\affiliation{Department of Physics, National Tsing Hua University, Hsinchu 30013, Taiwan}
\affiliation{School of Mathematics and Physics, The University of Queensland, St. Lucia, QLD 4072, Australia}
\affiliation{Centre for Modern Physics, Chongqing University, Chongqing 400044, The People's Republic of China}

\author{Murray T. Batchelor}
\affiliation{Australian National University, Canberra, ACT 2601, Australia}  
\affiliation{Centre for Modern Physics,
	Chongqing University, Chongqing 400044, The People's Republic of China}

\maketitle
\section{Ground state degeneracies}
\label{SMSectionA}

As remarked in the main text, the Hamiltonian possesses the staggered symmetry group ${\rm SU}(3)$ under both PBCs and  OBCs if the system size $L$ is even.
The Hamiltonian also enjoys additional symmetries, specifically time-reversal symmetry $K$, together with a $Z_2$ symmetry operation $\sigma$ -- the one-site translation operation $T_1$ under PBCs, and the bond-centred inversion operation $I_b$ under OBCs.
When $L$ is odd, the symmetry group is the same as that when $L$ is even under OBCs.
In contrast, there is a parity effect between even $L$'s and odd $L$'s under PBCs, for which the Hamiltonian does not possess the staggered symmetry group ${\rm SU}(3)$.
The symmetry group reduces to ${\rm SU(2)}$, generated by $S^x$, $S^y$ and $S^z$.
For simplicity, in this work we restrict ourselves to $L$ even, and leave a detailed discussion on the parity effect to a forthcoming publication.

Our claim is that all of the linearly independent ground states may be generated from the repeated action of the lowering operators $F_1$ and $F_2$ on the highest weight state as well as the generalized highest weight states, together with
the time-reversal operations $K$ and the additional $Z_2$ symmetry operation $\sigma$.
This justifies the necessity to introduce the generalized highest weight states.
In fact, although the degenerate ground states thus generated are not orthogonal to each other,
the number of linearly independent ground states equals the dimensionality of the ground state subspace.
Our methodology will be demonstrated for $L=2$, 4, and 6 below.
As will become clear, the usefulness of the generalized highest weight states lies in the fact that they allow us to uncover the self-similarities via an exact Schmidt decomposition, and  also make it possible to evaluate the entanglement entropy, leading to a logarithmic  scaling relation with the block size $n$.

To proceed, we introduce a generalized highest weight state, which is an unentangled (factorized) ground state.
In contrast to the highest weight state, the generalized highest states are not unique, and depend on the boundary conditions.
We remark that the staggered symmetry group ${\rm SU(3)}$ includes the uniform ${\rm SU(2)}$, generated by $S^x$, $S^y$ and $S^z$ as a subgroup.
Therefore, it is convenient to label both the highest weight state and the generalized highest weight states in terms of the eigenvalues of $S^z$.

Formally, the $m$-th generalized highest weight state $|{\rm hws}\rangle_g^m$ is defined recursively as \\
$|{\rm hws}\rangle_g^m\notin V_0\oplus V_1\oplus... V_{m-1}$, $E_1|{\rm hws}\rangle_g^m=0\mod(V_0\oplus V_1\oplus... V_{m-1})$, and $E_2|{\rm hws}\rangle_g^m=0\mod(V_0\oplus V_1\oplus... V_{m-1})$.
Here, $V_0$ denotes the subspace spanned by the degenerate ground states generated from the highest weight state, whereas $V_\mu$ denotes the subspace spanned by the degenerate ground states generated from the $\mu$-th generalized highest state, where  $\mu=1,2, \ldots, m$.
This means that, even though $|{\rm hws}\rangle_g^m$ are linearly independent to the states in the subspace  $V_0\oplus V_1\oplus... V_{m-1}$, all the states generated from $|{\rm hws}\rangle_g^m$, with $E_1^{M_1}E_2^{M_2}|{\rm hws}\rangle_g^m$ ($M_1\geq0$, $M_2\geq 0$ and $M_1+M_2\geq1$), are linearly dependent to the states in the subspace $V_0\oplus V_1\oplus... V_{m-1}$.
This definition works for any generalized highest weight states with $S^z$ eigenvalues greater than 0.
However, a generalized highest weight state, with $S^z$ eigenvalue 0, is peculiar, in the sense that for such a factorized ground state,
the action of the raising operators $E_1$ and $E_2$ does not yield any extra degenerate ground states, compared  to those generated from the action of the lowering operators $F_1$ and $F_2$, when the time-reversal operator $K$ and the additional $Z_2$ symmetry operator $\sigma$ are taken into account.

As an illustration, we consider the case when $L=4$.  We have 20 distinct factorized ground states under OBCs, in addition to the highest weight state  $|1111\rangle$ and the lowest weight state  $|\rm{-}1\rm{-}1\rm{-}1\rm{-}1\rangle$.
They take the form
\begin{align*}
	|0    \rm{-}1    \rm{-}1    \rm{-}1\rangle,\quad
	|\rm{-}1     0    \rm{-}1    \rm{-}1\rangle, \quad
	|1     0    \rm{-}1    \rm{-}1\rangle,\quad
	|\rm{-}1    \rm{-}1     0    \rm{-}1\rangle, \quad
    |0     1     0    \rm{-}1\rangle,\\
	|1     1     0    \rm{-}1\rangle, \quad
	|\rm{-}1    \rm{-}1    \rm{-}1     0\rangle,\quad
	|0    \rm{-}1    \rm{-}1     0\rangle, \quad 
	|\rm{-}1     0    \rm{-}1     0\rangle,\quad
	|1     0    \rm{-}1     0\rangle,\\
	|\rm{-}1     0     1     0\rangle,\quad
	|1     0     1     0\rangle, \quad
	|0     1     1     0\rangle,\quad
	|1     1     1     0\rangle,\quad
	|\rm{-}1    \rm{-}1     0     1\rangle,\\
	|0    \rm{-}1     0     1\rangle,\quad
	|1     1     0     1\rangle,\quad
	|\rm{-}1     0     1     1\rangle,\quad
	|1     0     1     1\rangle,\quad
	|0     1     1     1\rangle.
\end{align*}
It turns out that one may choose $|1110\rangle$ and $|0110\rangle$ as the generalized highest weight states.

In contrast, we have only 14 distinct factorized ground states under PBCs, in addition to the highest weight state $|1111\rangle$ and  the lowest weight state $|\rm{-}1\rm{-}1\rm{-}1\rm{-}1\rangle$.
They take the form
\begin{align*}
	|0    \rm{-}1    \rm{-}1    \rm{-}1\rangle, \quad
	|\rm{-}1     0    \rm{-}1    \rm{-}1\rangle,\quad
	|\rm{-}1    \rm{-}1     0    \rm{-}1\rangle,\quad
	|0     1     0    \rm{-}1\rangle,\quad 
	|\rm{-}1    \rm{-}1    \rm{-}1     0\rangle,\quad
	|\rm{-}1     0    \rm{-}1     0\rangle,\\
	|1     0   \rm{-}1     0\rangle,\quad
	|\rm{-}1     0     1     0\rangle,\quad
	|1     0     1     0\rangle,\quad
	|1     1     1     0\rangle,\quad
	|0    \rm{-}1     0     1\rangle,\quad
	|1     1     0     1\rangle,\quad
	|1     0     1     1\rangle, \quad
	|0     1     1     1\rangle.
\end{align*}
Here one may choose $|1110\rangle$ as the generalized highest weight state.

Now we demonstrate our methodology for $L=2$, 4 and 6 to see how the entire ground state subspace is spanned by the linearly independent degenerate ground states generated from the highest weight state as well as
the generalized highest weight states~\footnote{The explicit expressions for the linearly independent ground states are available upon request.}.

For $L=2$, there is no difference between OBCs and PBCs. There are 8 linearly independent ground states, generated from the repeated action of $F_1$ and $F_2$, combining with $K$ and $\sigma$, respectively, on the highest weight state $|{\rm hws}\rangle=|11\rangle$:
\begin{align*}
	|11\rangle,\quad
	F_1|11\rangle=|01\rangle,\quad
	\sigma F_1|11\rangle=|10\rangle,\quad
	F_2|11\rangle=|\rm{-}11\rangle-|1\rm{-}1\rangle,\\
	F_1\sigma F_1|11\rangle=|00\rangle+|\rm{-}11\rangle,\quad 
	F_1F_2|11\rangle=|0\rm{-}1\rangle,\quad
	\sigma F_1F_2|11\rangle=|\rm{-}10\rangle,\quad
	F_2^2|11\rangle=|\rm{-}1\rm{-}1\rangle.
\end{align*}

We remark that the action of the time-reversal operation $K$ on the above eight states does not yield any extra linearly independent ground states.

For $L=4$, 27 linearly independent ground states are generated from the repeated action of $F_1$ and $F_2$, combining with $K$ and $I_b$ under OBCs and with $K$ and $T_1$ under PBCs, on the highest weight state $|{\rm hws}\rangle=|1111\rangle$, with $S^z$ eigenvalue 4.
Accordingly, the subspace $V_0$ is spanned by the 27 linearly independent ground states, generated from the highest weight state $|1111\rangle$.
The factorized ground state $|1110\rangle$, with 
$S^z$ eigenvalue 3, turns out to be linearly independent to the states in the subspace $V_0$, and satisfies
$E_1|1110\rangle=0\mod(V_0)$ and $E_2|1110\rangle=0\mod(V_0)$.
Therefore, it is a generalized highest weight state, regardless of the boundary conditions.
Combining with $K$ and $I_b$ under OBCs and with $K$ and $T_1$ under PBCs, the repeated action of $F_1$ and $F_2$ on generalized highest weight state $|1110\rangle$, generates a sequence of degenerate ground states, with a total of 47 linearly independent ground states.
Hence there are an extra $20$ linearly independent ground states, which span the subspace $V_1$.
This exhausts the ground state subspace for PBCs, with $\Omega_L^{\rm PBC}=V_0\oplus V_1$, where ${\rm dim }(\Omega_L^{\rm PBC})=47$ for $L=4$.

Afterwards, for OBCs,  we may choose the factorized ground state $|0110\rangle$, with $S^z$ eigenvalue 2, as an extra generalized highest weight state.
The repeated action of $F_1$ and $F_2$, combining with $K$ and $I_b$, on $|0110\rangle$,  generates a sequence of degenerate ground states, with a total of 55 linearly independent ground states.
Hence there are an extra 8  linearly independent ground states spanning the subspace $V_2$.
This exhausts the ground state subspace for OBCs.
Here $\Omega_L^{\rm OBC}=V_0\oplus V_1\oplus V_2$, with ${\rm dim }(\Omega_L^{\rm OBC})=55$ for $L=4$.

Now we move to $L=6$.
The linearly independent ground states under PBCs may be constructed as follows.
In each case, we consider the repeated action of  $F_1$ and $F_2$, combining with $K$ and $T_1$.
First, the highest weight state $|111111\rangle$, with $S^z$ eigenvalue 6, generates 64 linearly independent ground states.
We denote the subspace spanned by the $64$ linearly independent ground states as $V_0$.
Second, the factorized ground state $|111110\rangle$, with $S^z$ eigenvalue 5, is a generalized highest weight state.
The repeated action on $|111110\rangle$ generates 204 linearly independent ground states.
Hence there are an extra 140  linearly independent ground states, which span the subspace $V_1$.
Third, the factorized ground state $|110110\rangle$, with $S^z$ eigenvalue 4, is an extra generalized highest weight state.
Repeated action on $|110110\rangle$ generates 285 linearly independent ground states.
Hence there are an extra 81 linearly independent ground states, which span the subspace $V_2$.
Fourth, it is found that all of the factorized ground states, with $S^z$ eigenvalues 3, 2, and 1, under PBCs, are all in the subspace $V_0\oplus V_1\oplus V_2$, respectively.
Therefore, we turn to the factorized ground states, with  $S^z$ eigenvalue 0.
It is peculiar, in the sense that the degenerate ground states, generated from the repeated action of $E_1$ and $E_2$, combining with $T_1$, on such a factorized ground state, may be identified with those  generated from the repeated action of $F_1$ and $F_2$, combining with $T_1$, on a factorized ground state of the {\it same} type, as long as the time-reversal operation $K$ is implemented.
That is, we only need to restrict to all of the extra degenerate ground states generated from the  repeated action of $F_1$ and $F_2$, combining with $K$ and $T_1$, on such a factorized ground state.
In other words, such a factorized ground state, with $S^z$  eigenvalue 0, may be recognized as a generalized highest weight state.
Repeated action on $|10\rm{-}1\rm{-}101\rangle$ generates 322 linearly independent ground states.
Hence there are an extra 37  linearly independent ground states, which span the subspace $V_3$.
This exhausts the ground state subspace for PBCs.
Here $\Omega_L^{\rm PBC}=V_0\oplus V_1\oplus V_2\oplus V_3$, with ${\rm dim }(\Omega_L^{\rm PBC})=322$ for $L=6$.

The linearly independent ground states under OBCs may be constructed as follows.
In each case, we consider the repeated action of $F_1$ and $F_2$, combining with $K$ and $I_b$. 
First, the highest weight state $|111111\rangle$, with $S^z$ eigenvalue 6, generates 64 linearly independent ground states. 
We denote the subspace spanned by the $64$ linearly independent ground states as the subspace $V_0$.
Second, the factorized ground state $|111110\rangle$,  with $S^z$ eigenvalue 5, is a generalized highest weight state.
Repeated action on $|111110\rangle$ generates 132 linearly independent ground states.
Hence there are an extra 68 linearly independent ground states, which span the subspace $V_1$.
Third, the factorized ground state $|111101\rangle$,  with $S^z$ eigenvalue 5, is an extra generalized highest weight state.
Repeated action on $|111101\rangle$ generates 202 linearly independent ground states.
Hence there are an extra 70 linearly independent ground states, which span the subspace $V_2$.
Fourth, the factorized ground state $|111011\rangle$, with $S^z$ eigenvalue 5, is an extra generalized highest weight state.
Repeated action on $|111011\rangle$ generates 204 linearly independent ground states.
Hence there are an extra 2  linearly independent ground states, which span the subspace $V_3$.
Fifth, the factorized ground state $|101101\rangle$,  with $S^z$ eigenvalue 4, is an extra generalized highest weight state.
Repeated  on $|101101\rangle$ generates 232 linearly independent ground states.
Hence there are extra 28  linearly independent ground states, which span the subspace $V_4$.
Sixth, the factorized ground state $|110110\rangle$,  with $S^z$ eigenvalue 4, is an extra generalized highest weight state.
Repeated action on $|110110\rangle$ generates 285 linearly independent ground states.
Hence there are an extra 53  linearly independent ground states, which span the subspace $V_5$.
Seventh, the factorized ground state $|011010\rangle$,  with $S^z$ eigenvalue 3, is an extra generalized highest weight state.
Repeated action on $|011010\rangle$ generates 322 linearly independent ground states.
Hence there are an extra 27 linearly independent ground states, which span the subspace $V_6$.
Eighth, the factorized ground state $|\rm{-}101101\rangle$,  with $S^z$ eigenvalue 2, is an extra generalized highest weight state.
Repeated action on $|\rm{-}101101\rangle$ generates 368 linearly independent ground states.
Hence there are an extra 46 linearly independent ground states, which span the subspace $V_7$.
Ninth, it is found that all of the factorized ground states, with $S^z$ eigenvalue 1, under OBCs, are all in the subspace $V_0\oplus V_1\oplus V_2\oplus V_3\oplus V_4\oplus V_5\oplus V_6\oplus V_7$.
Therefore, we turn to the factorized ground states, with  $S^z$ eigenvalue 0.
Here it is peculiar, in the sense that the degenerate ground states, generated from the repeated action of $E_1$ and $E_2$, combining with $I_b$, on such a factorized ground state, may be identified with those generated from the repeated action of $F_1$ and $F_2$, combining with $I_b$, on a factorized ground state of the {\it same} type, as long as the time-reversal operation $K$ is implemented.
In other words, we only need to restrict to all the extra degenerate ground states generated from the repeated action of $F_1$ and $F_2$, combining with $K$ and $I_b$, on such a factorized ground state.
That is, such a factorized ground state, with $S^z$ eigenvalue 0, may be recognized as a generalized highest weight state.
Therefore, the factorized ground state $|10\rm{-}1\rm{-}101\rangle$,  with $S^z$ eigenvalue 0, 
is an extra generalized highest weight state.
Repeated action on $|10\rm{-}1\rm{-}101\rangle$ generates 377 linearly independent ground states.
Hence there are  an extra 12 linearly independent ground states, which span the subspace $V_8$.
This exhausts the ground state subspace for OBCs.
So here $\Omega_L^{\rm OBC}=V_0\oplus V_1\oplus V_2\oplus V_3\oplus V_4\oplus V_5\oplus V_6\oplus V_7\oplus V_8$, with ${\rm dim }(\Omega_L^{\rm OBC})=377$ for $L=6$.

Our construction explains in detail why the ground state degeneracies $\Omega_L^{\rm OBC}$ and $\Omega_L^{\rm PBC}$ are different under OBCs and PBCs.

\section{Derivation of the normalization factors $Z_2(L,M_1,M_2)$, $Z_2(L,M_1,M_3)$, $Z_4(L,M_2,M_3)$ and $Z_6(L,M_1,M_2)$}
\label{AppD}

In this Section we give a derivation of the normalization factors $Z_2(L,M_1,M_2)$, $Z_2(L,M_1,M_3)$, $Z_4(L,M_2,M_3)$ and $Z_6(L,M_1,M_2)$ appearing in Eqs.~(7), (9), (11) and (13) for the highly degenerate ground states.
Specifically, we outline a derivation of the norms given in Eqs.~(8), (10), (12) and (14), with the period two, two, four and six, respectively.
This involves an extension of some technical tricks in combinatorics developed in Refs.~[27,28].
Although possible in principle, it is a formidable task to evaluate such norms for a degenerate ground state generated from the repeated action of the lowering operators $F_1$ and $F_3$, combined with the additional $Z_2$ symmetry operator $\sigma$, on a generic generalized highest weight state, if the period is increased up to the system size $L$.

(1) Consider first the degenerate ground states $|L,M_1,M_2\rangle_2$ in Eq.~(7). Due to the staggered nature of the symmetry group $\rm{SU(3)}$, 
we need to introduce a unit cell consisting of the two nearest-neighbour sites. 
There are therefore six possible configurations in the unit cell, namely  $|11\rangle$, $|1\rm{-}1\rangle$, $|01\rangle$, $|0\rm{-}1\rangle$, $|\rm{-}11\rangle$ and $|\rm{-}1\rm{-}1\rangle$.
The number of unit cells in these configurations are respectively labelled by $N_{11}$, $N_{1-1}$, $N_{01}$, $N_{0-1}$, $N_{-11}$ and $N_{-1-1}$.
Here $L$ must be a multiple of two.

The degenerate ground states $|L,M_1,M_2\rangle_2$ may then be rewritten as
	\begin{align}
	|L,M_1,M_2\rangle_2=&\frac{1}{Z_2(L,M_1,M_2)} \nonumber \\
&{\sum}'_{N_{11},N_{1\rm{-}1},N_{01}\atop N_{0-1},N_{\rm{-}11},N_{\rm{-}1\rm{-}1}}\!(-1)^{N_{13}+N_{23}+N_{33}}\!\sum_{P}\!
	|\underbrace{11...11}_{N_{11}}\!|\underbrace{1\rm{-}1...1\rm{-}1}_{N_{1\rm{-}1}}\!|\underbrace{01...01}_{N_{01}}\!|\underbrace{0\rm{-}1...0\rm{-}1}_{N_{0\rm{-}1}}
	\!|\underbrace{\rm{-}11...\rm{-}11}_{N_{\rm{-}11}}\!|\underbrace{\rm{-}1\rm{-}1...\rm{-}1\rm{-}1}_{N_{-1-1}}\rangle \,.
	\label{lm1m2D}
	\end{align}
The sum $\sum_{P}$ is taken over all permutations $P$ for a given partition $\{N_{11}$, $N_{1-1}$, $N_{01}$, $N_{0-1}$, $N_{-11}$, $N_{-1-1}\}$.
The sum ${\sum}'$ is over all possible values of $N_{11}$, $N_{1-1}$, $N_{01}$, $N_{0-1}$, $N_{-11}$ and $N_{-1-1}$, subject to the constraints $N_{01}+N_{0-1}=M_1$,  $N_{0-1}+2N_{-1-1}+N_{1-1}+N_{-11}=M_2$, and $N_{11}+N_{1-1}+N_{01}+N_{0-1}+N_{-11}+N_{-1-1}=L/2$.

The normalization factor $Z_2(L,M_1,M_2)$ takes the form
\begin{align}
	Z_2(L,M_1,M_2)=&M_1!M_2!\nonumber \\
&\sqrt{\!{\sum}'_{N_{11},N_{1-1},N_{01}\atop N_{0-1},N_{-11},N_{-1-1}} \! C_{L/2}^{N_{11}}C_{L/2-N_{11}}^{N_{1-1}}\!C_{L/2-N_{11}-N_{1-1}}^{N_{01}}\!C_{L/2-N_{11}-N_{1-1}-N_{01}}^{N_{0-1}}\!
		C_{L/2-N_{11}-N_{1-1}-N_{01}-N_{0-1}}^{N_{-11}}C_{L/2-N_{11}-N_{1-1}-N_{01}-N_{0-1}-N_{-11}}^{N_{-1-1}}}.
\label{zlm1m2}
\end{align}
It may be simplified as
\begin{equation}
Z_2(L,M_1,M_2)=M_1!M_2!\sqrt{C_{L/2}^{M_1}C_{L-M_1}^{M_2}} \,.
\label{zm1m2D}
\end{equation}
Provided $M_1\leq L/2$ and $M_2\leq L$, we have 
$H_1|L,M_1,M_2\rangle_2=(L/2-2M_1-M_2)|L,M_1,M_2\rangle_2$ and \newline $H_2|L,M_1,M_2\rangle_2=(L-M_1-2M_2)|L,M_1,M_2\rangle_2$.

(2) Now consider the degenerate ground states $|L,M_1,M_3\rangle_2$ in Eq.~(9), for which a unit cell consisting of the two nearest-neighbor sites again needs to be introduced, and $L$ must be a multiple of two.
The six possible configurations in the unit cell are $|11\rangle$, $|1\rm{-}1\rangle$, $|01\rangle$, $|0\rm{-}1\rangle$, $|10\rangle$  and $|00\rangle$.
The numbers of unit cells in these configurations are respectively labelled by $N_{11}$, $N_{1-1}$, $N_{01}$, $N_{0-1}$, $N_{10}$ and $N_{00}$. 

The degenerate ground states are written as 
	\begin{equation}
	|L,M_1,M_3\rangle_2=\frac{1}{Z_2(L,M_1,M_3)}{\sum}'_{N_{11},N_{1-1},N_{01}\atop N_{0-1},N_{10},N_{00}}\sum_{P}
	|\underbrace{11...11}_{N_{11}}|\underbrace{1-1...1-1}_{N_{1-1}}|\underbrace{01...01}_{N_{01}}|\underbrace{0-1...0-1}_{N_{0-1}}
	|\underbrace{10...10}_{N_{10}}|\underbrace{00...00}_{N_{00}}\rangle \,.
	\label{lm1m3D}
	\end{equation}
The sum $\sum_{P}$ is taken over all permutations $P$ for a given partition $\{N_{11}$, $N_{1-1}$, $N_{01}$, $N_{0-1}$, $N_{10}$, $N_{00}\}$.
The sum ${\sum}'$ is over all possible values of $N_{11}$, $N_{1-1}$, $N_{01}$, $N_{0-1}$, $N_{10}$ and $N_{00}$, subject to the constraints $N_{01}+N_{00}+N_{1-1}+2N_{0-1}=M_1$,  $N_{10}+N_{00}+N_{1-1}+N_{0-1}=M_3$, and $N_{11}+N_{1-1}+N_{01} +N_{0-1}+N_{10}+N_{00}=L/2$.
The normalization factor in Eq.~(\ref{lm1m3D}) takes the form
\begin{align}
	Z_2(L,M_1,M_3)=&M_1!M_3!\nonumber \\
&\sqrt{{\sum}'_{N_{11},N_{1-1},N_{01}\atop N_{0-1},N_{10},N_{00}} \!C_{L/2}^{N_{11}}\!C_{L/2-N_{11}}^{N_{1-1}}\!C_{L/2-N_{11}-N_{1-1}}^{N_{01}}\!C_{L/2-N_{11}-N_{1-1}-N_{01}}^{N_{0-1}}\!
		C_{L/2-N_{11}-N_{1-1}-N_{01}-N_{0-1}}^{N_{10}}\!C_{L/2-N_{11}-N_{1-1}-N_{01}-N_{0-1}-N_{-10}}^{N_{00}}}.
	\end{align}
It may be simplified as
\begin{equation}
Z_2(L,M_1,M_3)=M_1!M_3!\sqrt{C_{L/2}^{M_3}C_{L/2+M_3}^{M_1}}.
\label{zm1m3D}
\end{equation}
Provided $M_1\leq L/2+M_3$ and $M_3\leq L/2$, we have 
$H_1|L,M_1,M_3\rangle_2=(L/2-2M_1+M_3)|L,M_1,M_3\rangle_2$ and $H_2|L,M_1,M_3\rangle_2=(L-M_1-M_3)|L,M_1,M_3\rangle_2$.

(3) Now consider the degenerate ground states $|L,M_2,M_3\rangle_4$ in Eq.~(11).
The twelve possible configurations in a unit cell are $|1110\rangle$, $|\rm{-}1110\rangle$, $|1010\rangle$, $|\rm{-}1010\rangle$, $|1\rm{-}110\rangle$, $|\rm{-}1\rm{-}110\rangle$, $|11\rm{-}10\rangle$, $|\rm{-}11\rm{-}10\rangle$, $|10\rm{-}10\rangle$, $|\rm{-}10\rm{-}10\rangle$, $|1\rm{-}1\rm{-}10\rangle$ and  $|\rm{-}1\rm{-}1\rm{-}10\rangle$, consisting of the four adjacent sites.
The number of unit cells in these configurations are respectively labelled by $N_{1110}$, $N_{-1110}$, $N_{1010}$, $N_{-1010}$, $N_{1-110}$, $N_{-1-110}$,  $N_{11-10}$, $N_{-11-10}$, $N_{10-10}$, $N_{-10-10}$, $N_{1-1-10}$ and $N_{-1-1-10}$.
Here $L$ must be a multiple of four.

The degenerate ground states $|L,M_2,M_3\rangle_4$ take the form
\begin{align}
	|L,M_2,M_3\rangle_4=&\frac{1}{Z_4(L,M_2,M_3)}{\sum}'_{N_{1110},N_{-1110},N_{1010},
		N_{-1010},N_{1-110},N_{-1-110}\atop N_{11-10},N_{-11-10},N_{10-10},
		N_{-10-10},N_{1-1-10},N_{-1-1-10}} \nonumber \\
&\sum_{P}
	|\underbrace{1110...1110}_{N_{1110}}|\underbrace{\rm{-}1110...\rm{-}1110}_{N_{-1110}}|\underbrace{1010...1010}_{N_{1010}}|\underbrace{\rm{-}1010...\rm{-}1010}_{N_{-1010}}
	|\underbrace{1\rm{-}110...1\rm{-}110}_{N_{1-110}}\nonumber\\
&	|\underbrace{\rm{-}1\rm{-}110...\rm{-}1\rm{-}110}_{N_{-1-110}}	 |\underbrace{1110...1110}_{N_{1110}}|\underbrace{\rm{-}1110...\rm{-}1110}_{N_{-1110}}|\underbrace{1010...1010}_{N_{1010}}|\underbrace{\rm{-}1010...\rm{-}1010}_{N_{-1010}}\nonumber\\
	&|\underbrace{1\rm{-}110...1\rm{-}110}_{N_{1-110}}|\underbrace{\rm{-}1\rm{-}110...\rm{-}1\rm{-}110}_{N_{-1-110}}\rangle \,.
	\label{lm2m3Dexpand}
	\end{align}
The sum $\sum_{P}$ is over all permutations $P$ for a given partition
$\{N_{1110}$, $N_{-1110}$, $N_{1010}$, $N_{-1010}$, $N_{1-110}$, $N_{-1-110}$, $N_{11-10}$, $N_{-11-10}$, $N_{10-10}$,
$N_{-10-10}$, $N_{1-1-10}$, $N_{-1-1-10}\}$.
The sum ${\sum}'$ is taken over all possible values of
 $N_{1110}$, $N_{-1110}$, $N_{1010}$, $N_{-1010}$, $N_{1-110}$, $N_{-1-110}$, $N_{11-10}$, $N_{-11-10}$, $N_{10-10}$,
 $N_{-10-10}$, $N_{1-1-10}$, and $N_{-1-1-10}$,
subject to the constraints $N_{-1110}+N_{-1010}+N_{1-110}+2N_{-1-110}+ N_{11-10}+2N_{-11-10}+N_{10-10}+2N_{-10-10}+2N_{1-1-10}+3N_{-1-1-10}=M_2$,  $N_{1010}+N_{-1010}+N_{10-10}+N_{-10-10}=M_3$, and $N_{1110}+N_{-1110}+N_{1010}+N_{-1010}+N_{1-110}+N_{-1-110}+ N_{11-10}+N_{-11-10}+N_{10-10}+N_{-10-10}+N_{1-1-10}+N_{-1-1-10}=L/4$.
The normalization factor appearing in Eq.~(\ref{lm2m3Dexpand}) is given by
\begin{equation}
Z_4(L,M_2,M_3)=M_2!M_3!\sqrt{K(L,M_2,M_3)} \,,
\end{equation}
with
 	\begin{align}
	K(L,M_2,M_3)=&{\sum}'_{N_{1110},N_{-1110},N_{1010},
		N_{-1010},N_{1-110},N_{-1-110}\atop N_{11-10},N_{-11-10},N_{10-10},
		N_{-10-10},N_{1-1-10},N_{-1-1-10}} C_{L/4}^{N_{1110}}C_{L/4-N_{1110}}^{N_{-1110}}C_{L/4-N_{1110}-N_{-1110}}^{N_{1010}}	C_{L/4-N_{1110}-N_{-1110}-N_{1010}}^{N_{-1010}}\nonumber\\
	&		C_{L/4-N_{1110}-N_{-1110}-N_{1010}-N_{-1010}}^{N_{1-110}}
	C_{L/4-N_{1110}-N_{-1110}-N_{1010}-N_{-1010}-N_{1-110}}^{N_{-1-110}}
	C_{L/4-N_{1110}-N_{-1110}-N_{1010}-N_{-1010}-N_{1-110}-N_{-1-110}}^{N_{11-10}}\nonumber\\
	&
C_{L/4-N_{1110}-N_{-1110}-N_{1010}-N_{-1010}-N_{1-110}-N_{-1-110}-N_{11-10}}^{N_{-11-10}}C_{L/4-N_{1110}-N_{-1110}-N_{1010}-N_{-1010}-N_{1-110}-N_{-1-110}-N_{11-10}-N_{-11-10}}^{N_{10-10}}\nonumber\\
	&
	C_{L/4-N_{1110}-N_{-1110}-N_{1010}-N_{-1010}-N_{1-110}-N_{-1-110}-N_{11-10}-N_{-11-10}-N_{10-10}}^{N_{-10-10}}\nonumber\\
	&C_{L/4-N_{1110}-N_{-1110}-N_{1010}-N_{-1010}-N_{1-110}-N_{-1-110}-N_{11-10}-N_{-11-10}-N_{10-10}-N_{-10-10}}^{N_{1-1-10}}\nonumber\\	 &C_{L/4-N_{1110}-N_{-1110}-N_{1010}-N_{-1010}-N_{1-110}-N_{-1-110}-N_{11-10}-N_{-11-10}-N_{10-10}-N_{-10-10}-N_{1-1-10}}^{N_{-1-1-10}} \,.
	\end{align}
It may be simplified as
\begin{equation}
Z_4(L,M_2,M_3)=M_2!M_3!\sqrt{C_{L/4}^{M_3}C_{3L/4-M_3}^{M_2}}.
\label{zlm2m3g}
\end{equation}
Provided $M_2\leq 3L/4-M_3$ and $M_3\leq L/4$, 
we have $H_1|L,M_2,M_3\rangle_4=(3L/4-M_2+M_3)|L,M_2,M_3\rangle_4$ and \newline $H_2|L,M_2,M_3\rangle_4=(3L/4-2M_2-M_3)|L,M_2,M_3\rangle_4$.

(4) Turning to the degenerate ground states $|L,M_1,M_2\rangle_6$ in Eq.~(13), there are twelve possible configurations: $|10\rm{-}1\rm{-}101\rangle$, $|00\rm{-}1\rm{-}101\rangle$, $|\rm{-}10\rm{-}1\rm{-}101\rangle$, $|1\rm{-}1\rm{-}1\rm{-}101\rangle$, $|0\rm{-}1\rm{-}1\rm{-}101\rangle$, $|\rm{-}1\rm{-}1\rm{-}1\rm{-}101\rangle$,
$|10\rm{-}1\rm{-}10\rm{-}1\rangle$, $|00\rm{-}1\rm{-}10\rm{-}1\rangle$, $|\rm{-}10\rm{-}1\rm{-}10\rm{-}1\rangle$,\newline $|1\rm{-}1\rm{-}1\rm{-}10\rm{-}1\rangle$, $|0\rm{-}1\rm{-}1\rm{-}10\rm{-}1\rangle$, $|\rm{-}1\rm{-}1\rm{-}1\rm{-}10\rm{-}1\rangle$ in a unit cell, consisting of the six adjacent sites.
The number of unit cells in these configurations are respectively labelled by 
$N_{10-1-101}$, $N_{00-1-101}$, $N_{-10-1-101}$, $N_{1-1-1-101}$, $N_{0-1-1-101}$, $N_{-1-1-1-101}$, $N_{10-1-10-1}$, $N_{00-1-10-1}$, $N_{-10-1-10-1}$, $N_{1-1-1-10-1}$, $N_{0-1-1-10-1}$ and $N_{-1-1-1-10-1}$.
Here $L$ must be a multiple of six.

We may rewrite $|L,M_1,M_2\rangle_6$ as 
	\begin{align}
	|L,M_1,M_2\rangle_6=&\frac{1}{Z_6(L,M_1,M_2)}{\sum}'_{N_{10-1-101},N_{00-1-101},N_{-10-1-101}, N_{1-1-1-101},N_{0-1-1-101},N_{-1-1-1-101}\atop
		N_{10-1-10-1},N_{00-1-10-1},N_{-10-1-10-1}, N_{1-1-1-10-1},N_{0-1-1-10-1},N_{-1-1-1-10-1}}\sum_{P}
	|\underbrace{10\rm{-}1\rm{-}101...10\rm{-}1\rm{-}101}_{N_{10-1-101}}\nonumber\\
	&|\underbrace{00\rm{-}1\rm{-}101...00\rm{-}1\rm{-}101}_{N_{00-1-101}}
	|\underbrace{\rm{-}10\rm{-}1\rm{-}101...\rm{-}10\rm{-}1\rm{-}101}_{N_{-10-1-101}}
	|\underbrace{1\rm{-}1\rm{-}1\rm{-}101...1\rm{-}1\rm{-}1\rm{-}101}_{N_{1-1-1-101}}\nonumber\\
	&|\underbrace{0\rm{-}1\rm{-}1\rm{-}101...0\rm{-}1\rm{-}1\rm{-}101}_{N_{0-1-1-101}}
	|\underbrace{\rm{-}1\rm{-}1\rm{-}1\rm{-}101...\rm{-}1\rm{-}1\rm{-}1\rm{-}101}_{N_{-1-1-1-101}}
	|\underbrace{10\rm{-}1\rm{-}10\rm{-}1...10\rm{-}1\rm{-}10\rm{-}1}_{N_{10-1-10-1}}\nonumber\\
	&|\underbrace{00\rm{-}1\rm{-}10\rm{-}1...00\rm{-}1\rm{-}10\rm{-}1}_{N_{00-1-10-1}}
	|\underbrace{\rm{-}10\rm{-}1\rm{-}10\rm{-}1...\rm{-}10\rm{-}1\rm{-}10\rm{-}1}_{N_{-10-1-10-1}}
	|\underbrace{1\rm{-}1\rm{-}1\rm{-}10\rm{-}1...1\rm{-}1\rm{-}1\rm{-}10\rm{-}1}_{N_{1-1-1-10-1}}\nonumber\\
	&|\underbrace{0\rm{-}1\rm{-}1\rm{-}10\rm{-}1...0\rm{-}1\rm{-}1\rm{-}10\rm{-}1}_{N_{0-1-1-10-1}}
	|\underbrace{\rm{-}1\rm{-}1\rm{-}1\rm{-}10\rm{-}1...\rm{-}1\rm{-}1\rm{-}1\rm{-}10\rm{-}1}_{N_{-1-1-1-10-1}}\rangle \,.
	\label{lm2m3}
	\end{align}
The sum $\sum_{P}$ is taken over all permutations $P$ for a given partition $\{N_{10-1-101}$, $N_{00-1-101}$, $N_{-10-1-101}$, $N_{1-1-1-101}$,  $N_{0-1-1-101}$, $N_{-1-1-1-101}$, $N_{10-1-10-1}$, $N_{00-1-10-1}$, $N_{-10-1-10-1}$, $N_{1-1-1-10-1}$, $N_{0-1-1-10-1}$, $N_{-1-1-1-10-1}\}$.
The sum ${\sum}'$ is taken 
over all possible unit cell values subject to the constraints $N_{00-1-101}+ N_{1-1-1-101}+2N_{0-1-1-101}+N_{-1-1-1-101}+N_{00-1-10-1}+N_{1-1-1-10-1}+2N_{0-1-1-10-1}+N_{-1-1-1-10-1}=M_1$,  $N_{-10-1-101}+N_{-1-1-1-101}+N_{10-1-10-1}+N_{00-1-10-1}+N_{-10-1-10-1}+ N_{1-1-1-10-1}+N_{0-1-1-10-1}+2N_{-1-1-1-10-1}=M_2$, and $N_{10-1-101}+N_{00-1-101}+N_{-10-1-101}+ N_{1-1-1-101}$, $N_{0-1-1-101}+N_{-1-1-1-101}+
	N_{10-1-10-1}+N_{00-1-10-1}+N_{-10-1-10-1}+N_{1-1-1-10-1}+N_{0-1-1-10-1}+N_{-1-1-1-10-1}=L/6$.
The normalization factor is given by 
\begin{equation}
Z_6(L,M_1,M_2)=M_1!M_2!\sqrt{K(L,M_1,M_2)},
\end{equation}
with
\begin{align}
K(L,M_1,M_2)=&{\sum}'_{N_{10-1-101},N_{00-1-101},N_{-10-1-101}, N_{1-1-1-101},N_{0-1-1-101},N_{-1-1-1-101}\atop
		N_{10-1-10-1},N_{00-1-10-1},N_{-10-1-10-1}, N_{1-1-1-10-1},N_{0-1-1-10-1},N_{-1-1-1-10-1}}
	C_{L/6}^{N_{10-1-101}}C_{L/6-N_{10-1-101}}^{N_{00-1-101}}
	C_{L/6-N_{10-1-101}-N_{00-1-101}}^{N_{-10-1-101}}\nonumber\\
	&C_{L/6-N_{10-1-101}-N_{00-1-101}-N_{-10-1-101}}^{N_{1-1-1-101}}
	C_{L/6-N_{10-1-101}-N_{00-1-101}-N_{-10-1-101}-N_{1-1-1-101}}^{N_{0-1-1-101}}\nonumber\\
	&C_{L/6-N_{10-1-101}-N_{00-1-101}-N_{-10-1-101}-N_{1-1-1-101}-N_{0-1-1-101}}^{N_{-1-1-1-101}}\nonumber\\
	&		C_{L/6-N_{10-1-101}-N_{00-1-101}-N_{-10-1-101}-N_{1-1-1-101}-N_{0-1-1-101}-N_{-1-1-1-101}}^{N_{10-1-10-1}}\nonumber\\
	&		C_{L/6-N_{10-1-101}-N_{00-1-101}-N_{-10-1-101}-N_{1-1-1-101}-N_{0-1-1-101}-N_{-1-1-1-101}-N_{10-1-10-1}}^{N_{00-1-10-1}} \\
	&		C_{L/6-N_{10-1-101}-N_{00-1-101}-N_{-10-1-101}-N_{1-1-1-101}-N_{0-1-1-101}-N_{-1-1-1-101}-N_{10-1-10-1}-N_{00-1-10-1}}^{N_{-10-1-10-1}}\nonumber\\
	&		C_{L/6-N_{10-1-101}-N_{00-1-101}-N_{-10-1-101}-N_{1-1-1-101}-N_{0-1-1-101}-N_{-1-1-1-101}-N_{10-1-10-1}-N_{00-1-10-1}-N_{-10-1-10-1}}^{N_{1-1-1-10-1}}\nonumber\\
	&		C_{L/6-N_{10-1-101}-N_{00-1-101}-N_{-10-1-101}-N_{1-1-1-101}-N_{0-1-1-101}-N_{-1-1-1-101}-N_{10-1-10-1}-N_{00-1-10-1}-N_{-10-1-10-1}-N_{1-1-1-10-1}}^{N_{0-1-1-10-1}}\nonumber\\
	&			C_{L/6-N_{10-1-101}-N_{00-1-101}-N_{-10-1-101}-N_{1-1-1-101}-N_{0-1-1-101}-N_{-1-1-1-101}-N_{10-1-10-1}-N_{00-1-10-1}-N_{-10-1-10-1}-N_{1-1-1-10-1}-N_{0-1-1-10-1}}^{N_{-1-1-1-10-1}} \,. \nonumber
	\end{align}
It may be simplified as
\begin{equation}
Z_6(L,M_1,M_2)=M_1!M_2!\sqrt{\sum_{l=0}^{M_1}C_{L/6}^{l}C_{L/6}^{M_1-l}C_{L/3-l}^{M_2}}.
\label{zlm1m2g}
\end{equation}
Provided $M_1\leq L/3$, $M_2\leq L/3$ and $M_1+2M_2\leq 5L/6$, 
we have $H_1|L,M_1,M_2\rangle_6=(-2M_1-M_2)|L,M_1,M_2\rangle_6$ and $H_2|L,M_1,M_2\rangle_6=(-M_1-2M_2)|L,M_1,M_2\rangle_6$. 

\section{A derivation of $Z_2(L,M_1,M_3)$ and $Z_2(L,M_2,M_3)$ for the highly degenerate ground states from the highest weight state without the one-site translational invariance}
\label{AppE}

For the ferromagnetic spin-1 biquadratic model under consideration, one may choose the highest weight ground state as $|\rm{hws}\rangle=|0_x0_y...0_x0_y\rangle$.
Accordingly, the eight generators for the staggered symmetry group $\rm{SU(3)}$  are the Cartan generators  $H_1$ and $H_2$, with the raising operators $E_1$, $E_2$, $E_3$ and the lowering operators $F_1$, $F_2$, $F_3$, as presented in Appendix A, satisfying $[H_\alpha,E_\alpha]=2E_\alpha$, $[H_\alpha,F_\alpha]=-2F_\alpha$, $[E_\alpha,F_\alpha]=H_\alpha$.
Since $[F_1,F_2]=0$ and $[F_2,F_3]=0$, two possible ways may be employed to generate sequences of the degenerate ground states. 
Firstly, a sequence of the degenerate ground states $|L,M_1,M_2\rangle_2=1/Z_2(L,M_1,M_2)F_1^{M_1}F_2^{M_2}|0_x0_y...0_x0_y\rangle$ ($M_1=0$, \ldots, $L/2$, $M_2=0$, \ldots, $L$) are generated from the repeated action of the lowering operators $F_1$ and $F_2$ on the highest weight state $|0_x0_y...0_x0_y\rangle$. 
Secondly, a sequence of the degenerate ground states $|L,M_2,M_3\rangle_2=1/Z_2(L,M_2,M_3)F_2^{M_2}F_3^{M_3}|0_x0_y...0_x0_y\rangle$ ($M_2=0$, \ldots, $L$, $M_3=0$, \ldots, $L/2$)  are generated from the repeated action of the lowering operators $F_2$ and $F_3$ on the highest weight state $|0_x0_y...0_x0_y\rangle$.
We remark that $L$ must be a multiple of two. 

	(1) $|L,M_1,M_2\rangle_2$ may be rewritten as 
	\begin{equation}
	|L,M_1,M_2\rangle_2=\frac{1}{Z_2(L,M_1,M_2)}\sum_{P}{\sum}'_{N_{xz},N_{yz}, N_{xx}\atop N_{yy},N_{yx}, N_{xy}}(-1)^{N_{yz}+N_{yy}+N_{yx}}
	i^{L/2+N_{yz}+N_{yx}-N_{xy}}|\underbrace{0_x0_z...}_{N_{xz}}|\underbrace{0_y0_z...}_{N_{yz}}|\underbrace{0_x0_x...}_{N_{xx}}
	|\underbrace{0_y0_y...}_{N_{yy}}|\underbrace{0_y0_x...}_{N_{yx}}|\underbrace{0_x0_y...}_{N_{xy}}\rangle \,.\label{lm1m2E}
	\end{equation}
The sum $\sum_{P}$ is taken over all permutations $P$ for a given partition $\{N_{xz}$, $N_{yz}$, $N_{xx}$, $N_{yy}$, $N_{yx}$, $N_{xy}\}$.
Here $N_{xz}$, $N_{yz}$, $N_{xx}$, $N_{yy}$, $N_{yx}$ and $N_{xy}$ denote the numbers of unit cells in the configurations $|0_x0_z\rangle$, $|0_y0_z\rangle$, $|0_x0_x\rangle$, $|0_y0_y\rangle$, $|0_y0_z\rangle$, and $|0_x0_y\rangle$, respectively.
The sum ${\sum}'$ is taken over all possible values of $N_{xz}$, $N_{yz}$, $N_{xx}$, $N_{yy}$, $N_{yx}$ and $N_{xy}$, subject to the constraints $N_{xz}+N_{yz}=M_1$, $N_{yz}+N_{xx}+N_{yy}+2N_{yx}=M_2$ and $N_{xz}+N_{yz}+N_{xx}+N_{yy}+N_{yx}+N_{xy}=L/2$.

The normalization factor $Z_2(L,M_1,M_2)$ takes the form
	\begin{equation} Z_2(L,M_1,M_2)=M_1!M_2!\sqrt{\sum_{N_{xz},N_{xx},N_{yx}}C_{L/2}^{N_{xz}}C_{L/2-N_{xz}}^{N_{yz}}C_{L/2-N_{xz}-N_{yz}}^{N_{xx}}C_{L/2-N_{xz}-N_{yz}-N_{xx}}^{N_{yy}}
		C_{L/2-N_{xz}-N_{yz}-N_{xx}-N_{yy}}^{N_{yx}}C_{L/2-N_{xz}-N_{yz}-N_{xx}-N_{yy}-N_{yx}}^{N_{xy}}}.
	\end{equation}
It may be simplified as
	\begin{align}
	Z_2(L,M_1,M_2)=M_1!M_2!\sqrt{C_{L/2}^{M_1}C_{L-M_1}^{M_2}}.
\label{zlm1m2E}
	\end{align}
Provided $M_1\leq L/2$ and $M_2\leq L$, 
we have $H_1|L,M_1,M_2\rangle_2=(L/2-2M_1-M_2)|L,M_1,M_2\rangle_2$ and $H_2|L,M_1,M_2\rangle_2=(L-M_1-2M_2)|L,M_1,M_2\rangle_2$.
	
(2) $|L,M_2,M_3\rangle_2$ may be rewritten as 
	\begin{equation}
	|L,M_2,M_3\rangle_2=\frac{1}{Z_2(L,M_2,M_3)}\sum_{p}{\sum}'_{N_{zy},N_{zx},N_{xx}\atop N_{yy},N_{yx}, N_{xy}}(-1)^{M_3+N_{yy}+N_{yx}}i^{M_2}
	|\underbrace{0_z0_y...}_{N_{zy}}|\underbrace{0_z0_x...}_{N_{zx}}|\underbrace{0_x0_x...}_{N_{xx}}|\underbrace{0_y0_y...}_{N_{yy}}
	|\underbrace{0_y0_x...}_{N_{yx}}|\underbrace{0_x0_y...}_{N_{xy}}\rangle \,.\label{lm2m3E}
	\end{equation}
The sum $\sum_{P}$ is taken over all permutations $P$ for a given partition $\{N_{zy}$, $N_{zx}$, $N_{xx}$, $N_{yy}$, $N_{yx}$, $N_{xy}\}$.
Here $N_{zy}$, $N_{zx}$, $N_{xx}$, $N_{yy}$, $N_{yx}$ and $N_{xy}$ respectively denote the numbers of unit cells in the configurations $|0_z0_y\rangle$, $|0_z0_x\rangle$, $|0_x0_x\rangle$,  $|0_y0_y\rangle$, $|0_y0_z\rangle$ and $|0_x0_y\rangle$. 
The sum ${\sum}'$ is taken over all possible values of $N_{zy},N_{zx}, N_{xx},N_{yy},N_{yx}$ and $N_{xy}$ subject to the constraints $N_{zx}+N_{zy}=M_3$, $N_{yy}+N_{zx}+N_{xx}+2N_{yx}=M_2$ and $N_{zy}+ N_{zx}+N_{xx}+N_{yy}+N_{yx}+N_{xy}=L/2$.

The normalization factor is
	\begin{equation}
Z_2(L,M_2,M_3)=M_2!M_3!\sqrt{\sum_{N_{zy},N_{xx},N_{yx}}C_{L/2}^{N_{zy}}C_{L/2-N_{zy}}^{N_{zx}}C_{L/2-N_{zx}-N_{zy}}^{N_{xx}}C_{L/2-N_{zy}-N_{zx}-N_{xx}}^{N_{yy}}
		C_{L/2-N_{zy}-N_{zx}-N_{xx}-N_{yy}}^{N_{yx}}C_{L/2-N_{zy}-N_{zx}-N_{xx}-N_{yy}-N_{yx}}^{N_{xy}}}.
	\end{equation}
It may be simplified as
	\begin{align}
	Z_2(L,M_2,M_3)=M_2!M_3!\sqrt{C_{L/2}^{M_3}C_{L-M_3}^{M_2}}.
\label{zlm2m3n}
	\end{align}
Provided $M_2\leq L$ and $M_3\leq L/2$, we have 
$H_1|L,M_2,M_3\rangle_2=(L/2-M_2+M_3)|L,M_2,M_3\rangle_2$ and $H_2|L,M_2,M_3\rangle_2=(L-2M_2-M_3)|L,M_2,M_3\rangle_2$.

\section{Schmidt decomposition and entanglement entropy}

In this Appendix we outline a way to evaluate the entanglement entropy for the degenerate ground states (i) $|L,M_1,M_2\rangle_2$, (ii) $|L,M_1,M_3\rangle_2$,  (iii) $|L,M_2,M_3\rangle_4$, and (iv) $|L,M_1,M_2\rangle_6$ appearing in Eqs.~(7), (9), (11) and (13) of the main text. 

(i) The states $|L,M_1,M_2\rangle_2$ in Eq.~(7) admit the exact Schmidt decomposition
\begin{equation}
|L,M_1,M_2\rangle_2= \sum\limits_{k_1=0}^{\min(M_1,n/2)}\sum\limits_{k_2=0}^{\min(M_2,n-k_1)}\lambda(L,n,k_1,k_2,M_1,M_2)
|n,k_1,k_2\rangle_2|L-n,M_1-k_1,M_2-k_2\rangle_2 \,. \label{svd}
\end{equation}
Here $n$ is even, and the Schmidt coefficients $\lambda(L,n,k_1,k_2,M_1,M_2)$ are
\begin{equation}
\lambda(L,n,k_1,k_2,M_1,M_2)=C_{M_1}^{k_1}C_{M_2}^{k_2}\frac{Z_2(n,k_1,k_2)Z_2(L-n,M_1-k_1,M_2-k_2)}{Z_2(L,M_1,M_2)}.
\label{m1m2}
\end{equation}
The three $Z_2$ terms take the same form as $Z_2(L,M_1,M_2)$ in Eq.~(8).
It follows that $\lambda(L,n,k_1,k_2,M_1,M_2)$ can be rewritten as
\begin{equation}
\lambda(L,n,k_1,k_2,M_1,M_2)=\sqrt{\frac{C_{n/2}^{k_1}C_{n-k_1}^{k_2}C_{(L-n)/2}^{M_1-k_1}C_{L-n-M_1+k_1}^{M_2-k_2}} {C_{L/2}^{M_1}C_{L-M_1}^{M_2}}} \,.
\label{lamm1m2}
\end{equation}
The entanglement entropy $S_{\!L}(n,M_1,M_2)$ for 
the degenerate ground states $|L,M_1,M_2\rangle_2$ then  follows from
\begin{align}
S_{\!L}(n,M_1,M_2)=-\sum\limits_{k_1=0}^{\min(M_1,n/2)}\sum\limits_{k_2=0}^{\min(M_2,n-k_1)} \Lambda(L,n,k_1,k_2,M_1,M_2)
\log_{2}\Lambda(L,n,k_1,k_2,M_1,M_2),
\label{snkm1m2}
\end{align}
where $\Lambda(L,n,k_1,k_2,M_1,M_2)$ are the eigenvalues of the reduced density matrix $\rho_L(n,M_1,M_2)$, with \\ $\Lambda(L,n,k_1,k_2,M_1,M_2)=[\lambda(L,n,k_1,k_2,M_1,M_2)]^2$.

In the thermodynamic limit $L \rightarrow \infty$, with 
fixed fillings $f_1=M_1/L$ and $f_2=M_2/L$, the eigenvalues of the reduced density matrix $\Lambda(L,n,k_1,k_2,M_1,M_2)$ in Eq.~(\ref{snkm1m2}) become
\begin{equation}
\Lambda_\infty(n,k_1,k_2,f_1,f_2)=C_{n-k_1}^{k_2}v^{k_2}(1-v)^{n-k_1-k_2}C_{n/2}^{k_1}u^{k_1}(1-u)^{n/2-k_1},
\end{equation}
where $u=2f_1$ and $v=f_2/(1-f_1)$.
In the thermodynamic limit, the entanglement entropy $S(n,f_1,f_2)$ thus follows from
\begin{align}
S(n,f_1,f_2)=-\sum\limits_{k_1=0}^{\min(M_1,n/2)}\sum\limits_{k_2=0}^{\min(M_2,n-k_1)}
\Lambda_\infty(n,k_1,k_2,f_1,f_2)
\log_{2}\Lambda_\infty(n,k_1,k_2,f_1,f_2).
\label{snkf1}
\end{align}

(ii) The states $|L,M_1,M_3\rangle_2$ in Eq.~(9) admit the exact Schmidt decomposition
\begin{equation}
|L,M_1,M_3\rangle_2=\sum\limits_{k_1=0}^{\min(M_1,n/2+k_3)}\sum\limits_{k_3=0}^{\min(M_3,n/2)} 
|n,k_1,k_3\rangle_2|L-n,M_1-k_1,M_3-k_3\rangle_2 \,. \label{svd}
\end{equation}
Here $n$ is even, and the Schmidt coefficients $\lambda(L,n,k_1,k_3,M_1,M_3)$ are
\begin{equation}
\lambda(L,n,k_1,k_3,M_1,M_3)=C_{M_1}^{k_1}C_{M_3}^{k_3}\frac{Z_2(n,k_1,k_3)Z_2(L-n,M_1-k_1,M_3-k_3)}{Z_2(L,M_1,M_3)}.
\label{m1m3}
\end{equation}
The three $Z_2$ terms take the same form as $Z_2(L,M_1,M_3)$ in Eq.~(10).
It follows that $\lambda(L,n,k_1,k_3,M_1,M_3)$ can be rewritten as
\begin{equation}
\lambda(L,n,k_1,k_3,M_1,M_3)=\sqrt{\frac{C_{n/2}^{k_3}C_{n/2+k_3}^{k_1}C_{(L-n)/2}^{M_3-k_3}C_{(L-n)/2+M_3-k_3}^{M_1-k_1}} {C_{L/2}^{M_3}C_{L/2+M_3}^{M_1}}}.
\label{lamm1m3}
\end{equation}
The entanglement entropy $S_{\!L}(n,M_1,M_3)$ for 
the degenerate ground states $|L,M_1,M_3\rangle_2$ then follows from
\begin{align}
S_{\!L}(n,M_1,M_3)=-\sum\limits_{k_1=0}^{\min(M_1,n/2+k_3)}\sum\limits_{k_3=0}^{\min(M_3,n/2)} \Lambda(L,n,k_1,k_3,M_1,M_3)
\log_{2}\Lambda(L,n,k_1,k_3,M_1,M_3),
\label{snkm1m3}
\end{align}
where $\Lambda(L,n,k_1,k_3,M_1,M_3)$ are the eigenvalues of the reduced density matrix $\rho_L(n,M_1,M_3)$, with \\ $\Lambda(L,n,k_1,k_3,M_1,M_3)=[\lambda(L,n,k_1,k_3,M_1,M_3)]^2$.

In the thermodynamic limit $L \rightarrow \infty$, 
with fixed fillings $f_1=M_1/L$ and $f_3=M_3/L$, the eigenvalues of the reduced density matrix $\Lambda(L,n,k_1,k_3,M_1,M_3)$ in Eq.~(\ref{snkm1m3}) become
\begin{equation}
\Lambda_\infty(n,k_1,k_3,f_1,f_3)=C_{n/2+k_3}^{k_1}v^{k_1}(1-v)^{n/2+k_3-k_1}C_{n/2}^{k_3}u^{k_3}(1-u)^{n/2-k_3},
\end{equation}
where $u=2f_3$ and $v=2f_1/(1+2f_3)$.
In the thermodynamic limit, the entanglement entropy $S(n,f_1,f_3)$ is thus 
\begin{align}
S(n,f_1,f_3)=-\sum\limits_{k_1=0}^{\min(M_1,n/2+k_3)}\sum\limits_{k_3=0}^{\min(M_3,n/2)} 
\Lambda_\infty(n,k_1,k_3,f_1,f_3)
\log_{2}\Lambda_\infty(n,k_1,k_3,f_1,f_3).
\label{snkf3f1}
\end{align}

(iii) The states $|L,M_2,M_3\rangle_4$ in Eq.~(11) admit the exact Schmidt  decomposition
\begin{equation}
|L,M_2,M_3\rangle_4= \sum\limits_{k_2=0}^{\min(M_2,3n/4-k_3)}\sum\limits_{k_3=0}^{\min(M_3,n/4)} \lambda(L,n,k_2,k_3,M_2,M_3)
|n,k_2,k_3\rangle_4|L-n,M_2-k_2,M_3-k_3\rangle_4 \,. \label{svd}
\end{equation}
Here $n$ is a multiple of four, and the Schmidt coefficients 
 $\lambda(L,n,k_2,k_3,M_2,M_3)$ are
\begin{equation}
\lambda(L,n,k_2,k_3,M_2,M_3)=C_{M_2}^{k_2}C_{M_3}^{k_3}\frac{Z_4(n,k_2,k_3)Z_4(L-n,M_2-k_2,M_3-k_3)}{Z_4(L,M_2,M_3)}.
\label{m2m3}
\end{equation}
The three $Z_4$ terms take the same form as $Z_4(L,M_2,M_3)$ in Eq.~(\ref{zlm2m3g}).
So the above equation can be rewritten as
\begin{equation}
\lambda(L,n,k_2,k_3,M_2,M_3)=\sqrt{\frac{C_{n/4}^{k_3}C_{3n/4-k_3}^{k_2}C_{(L-n)/4}^{M_3-k_3}C_{3(L-n)/4-M_3+k_3}^{M_2-k_2}} {C_{L/4}^{M_3}C_{3L/4-M_3}^{M_2}}}.
\label{lamm2m3}
\end{equation}
The entanglement entropy $S_{\!L}(n,M_2,M_3)$ for the degenerate ground states $|L,M_1,M_3\rangle_4$ is then
\begin{align}
S_{\!L}(n,M_2,M_3)=-\sum\limits_{k_2=0}^{\min(M_2,3n/4-k_3)}\sum\limits_{k_3=0}^{\min(M_3,n/4)}\Lambda(L,n,k_2,k_3,M_2,M_3)
\log_{2}\Lambda(L,n,k_2,k_3,M_2,M_3) \,,
\label{snkm2m3}
\end{align}
where $\Lambda(L,n,k_2,k_3,M_2,M_3)$ are the eigenvalues of the reduced density matrix $\rho_L(n,M_2,M_3)$, with\newline $\Lambda(L,n,k_2,k_3,M_2,M_3)=[\lambda(L,n,k_2,k_3,M_2,M_3)]^2$.

In the thermodynamic limit $L \rightarrow \infty$, 
with fixed fillings $f_2=M_2/L$ and $f_3=M_3/L$, the eigenvalues of the reduced density matrix in Eq.~(\ref{snkm2m3}) become
\begin{equation}
\Lambda_\infty(n,k_2,k_3,f_2,f_3)=C_{3n/4-k_3}^{k_2}v^{k_2}(1-v)^{3n/4-k_3-k_2}C_{n/4}^{k_3}u^{k_3}(1-u)^{n/4-k_3},
\end{equation}
where $u=4f_3$ and $v=4f_2/(3-4f_3)$.
In the thermodynamic limit, the entanglement entropy $S(n,f_2,f_3)$ is then 
\begin{align}
S(n,f_2,f_3)=-\sum\limits_{k_2=0}^{\min(M_2,3n/4-k_3)}\sum\limits_{k_3=0}^{\min(M_3,n/4)}
\Lambda_\infty(n,k_2,k_3,f_2,f_3)
\log_{2}\Lambda_\infty(n,k_2,k_3,f_2,f_3).
\label{snkf2f3}
\end{align}

(iv) The states $|L,M_1,M_2\rangle_6$ in Eq.~(13) admit the exact Schmidt  decomposition
\begin{equation}
|L,M_1,M_2\rangle_6= \sum\limits_{k_1=0}^{\min(M_1,n/3)}\sum\limits_{k_2=0}^{\min(M_2,n/2-k_1)}\lambda(L,n,k_1,k_2,M_1,M_2)
|n,k_1,k_2\rangle_6|L-n,M_1-k_1,M_2-k_2\rangle_6, \label{svd}
\end{equation}
Here $n$ is a multiple of six, and the Schmidt coefficients  $\lambda(L,n,k_1,k_2,M_1,M_2)$ are
\begin{equation}
\lambda(L,n,k_1,k_2,M_1,M_2)=C_{M_1}^{k_1}C_{M_2}^{k_2}\frac{Z_6(n,k_1,k_2)Z_6(L-n,M_1-k_1,M_2-k_2)}{Z_6(L,M_1,M_2)}.
\label{m1m2g}
\end{equation}
The three $Z_6$ terms take the same form as $Z_6(L,M_1,M_2)$ in Eq.~(14).
It follows that 
\begin{equation}
\lambda(L,n,k_1,k_2,M_1,M_2)=
\frac{\sqrt{\sum_{p=0}^{k_1}C_{n/6}^{p}C_{n/6}^{k_1-p}C_{n/3-p}^{k_2}
\sum_{q=0}^{M_1-k_1}C_{(L-n)/6}^{q}C_{(L-n)/6}^{M_1-k_1-q}C_{(L-n)/3-q}^{M_2-k_2}}}
{\sqrt{\sum_{l=0}^{M_1}C_{L/6}^{l}C_{L/6}^{M_1-l}C_{L/3-l}^{M_2}}}.
\label{lamm1m2g}
\end{equation}
The entanglement entropy $S_{\!L}(n,M_1,M_2)$ for the degenerate ground states $|L,M_1,M_2\rangle_6$ is then
 \begin{align}
S_{\!L}(n,M_1,M_2)=-\sum\limits_{k_1=0}^{\min(M_1,n/3)}\sum\limits_{k_2=0}^{\min(M_2,n/2-k_1)}\Lambda(L,n,k_1,k_2,M_1,M_2)
\log_{2}\Lambda(L,n,k_1,k_2,M_1,M_2),
\label{snkm1m2g}
\end{align}
where $\Lambda(L,n,k_1,k_2,M_1,M_2)$ are the eigenvalues of the reduced density matrix $\rho_L(n,M_1,M_2)$, with \\
$\Lambda(L,n,k_1,k_2,M_1,M_2)=[\lambda(L,n,k_1,k_2,M_1,M_2)]^2$.

In the thermodynamic limit $L \rightarrow \infty$, 
with fixed fillings $f_1=M_1/L$ and $f_2=M_2/L$ where $f_2\leq1/6$, the eigenvalues of the reduced density matrix appearing in Eq.~(\ref{snkm1m2g}) become
\begin{equation}
\Lambda_\infty(n,k_1,k_2,f_1,f_2)=\sum_{l=0}^{\min(k_1,n/6)}C_{n/6}^{l}C_{n/6}^{k_1-l}u^{k_1}(1-u)^{n/3-k_1}
C_{n/3-l}^{k_2}v^{k_2}(1-v)^{n/3-l-k_2},
\end{equation}
where $u=3f_1$ and $v=3f_2/(1-3l/n)$.
In the thermodynamic limit the entanglement entropy $S(n,f_1,f_2)$ is then 
\begin{align}
S(n,f_1,f_2)=-\sum\limits_{k_1=0}^{\min(M_1,n/3)}\sum\limits_{k_2=0}^{\min(M_2,n/2-k_1)}
\Lambda_\infty(n,k_1,k_2,f_1,f_2)
\log_{2}\Lambda_\infty(n,k_1,k_2,f_1,f_2).
\label{snkf1f2}
\end{align}